\documentclass[useAMS,usenatbib,usegraphicx]{mn2e}

\newcommand{\kms}{\textrm{km~s$^{-1}$}}

\newcommand{\ergs}{\textrm{erg~s$^{-1}$}}
\newcommand{\lsolar}{L$_{\odot}$}
\newcommand{\msolar}{M$_{\odot}$}
\newcommand{\ml}{M$_{\odot}$ yr$^{-1}$}
\newcommand{\mdot}{\dot{M}}

\PassOptionsToPackage{normalem}{ulem}
\usepackage{ulem}
\usepackage{xcolor}

\usepackage{subfigure}
\usepackage{multirow}
\usepackage{amssymb}

\title[The 3000-Day Light Curve of SN 2006gy]{What Powers the 3000-Day Light Curve of SN 2006gy?}
\author[O. D. Fox et al.]{Ori D. Fox$^{1,2,3}$, Nathan Smith$^{4}$, S. Mark Ammons$^{5}$, Jennifer Andrews$^{4}$, \newauthor K. Azalee Bostroem$^{6}$, S. Bradley Cenko$^{7,8}$, Geoffrey C. Clayton$^{9}$, Eli Dwek$^{7}$, \newauthor Alexei V. Filippenko$^{1}$, Joseph S. Gallagher$^{10}$, Patrick L. Kelly$^{1}$, Jon C. Mauerhan$^{1}$, \newauthor Adam M. Miller$^{11,12,13}$, and Schuyler D. Van Dyk$^{14}$\\
$^{1}$Department of Astronomy, University of California, Berkeley, CA 94720-3411.\\
$^{2}$Space Telescope Science Institute, 3700 San Martin Drive, Baltimore, MD 21218, USA.\\
$^{3}$ofox@stsci.edu.\\
$^{4}$Steward Observatory, 933 N. Cherry Ave., Tucson, AZ 85721, USA.\\
$^{5}$Lawrence Livermore National Laboratory, L-210 7000 East Ave, Livermore, CA 94550.\\
$^{6}$Department of Physics, University of California, Davis, CA 95616, USA.\\
$^{7}$Astrophysics Science Division, NASA Goddard Space Flight Center, Mail Code 661, Greenbelt, MD 20771, USA.\\
$^{8}$Joint Space-Science Institute, University of Maryland, College Park, MD 20742, USA.\\
$^{9}$Dept.\ of Physics \& Astronomy, Louisiana State University, Baton Rouge, LA 70803, USA.\\
$^{10}$University of Cincinnati Blue Ash College, 9555 Plainfield Rd., Blue Ash, OH 45236, USA.\\
$^{11}$Jet Propulsion Laboratory, 4800 Oak Grove Drive, MS 169-506, Pasadena, CA 91109, USA.\\
$^{12}$California Institute of Technology, Pasadena, CA 91125, USA.\\
$^{13}$Hubble Fellow.\\
$^{14}$IPAC/Caltech, Mailcode 100-22, Pasadena, CA 91125, USA}
\begin{document}

\maketitle
\begin{abstract}

SN 2006gy was the most luminous SN ever observed at the time of its discovery and the first of the newly defined class of superluminous supernovae (SLSNe).  The extraordinary energetics of SN 2006gy and all SLSNe ($>10^{51}$~erg) require either atypically large explosion energies (e.g., pair-instability explosion) or the efficient conversion of kinetic into radiative energy (e.g., shock interaction).  The mass-loss characteristics can therefore offer important clues regarding the progenitor system.  For the case of SN 2006gy, both a scattered and thermal light echo from circumstellar material (CSM) have been reported at later epochs (day $\sim800$), ruling out the likelihood of a pair-instability event and leading to constraints on the characteristics of the CSM.  Owing to the proximity of the SN to the bright host-galaxy nucleus, continued monitoring of the light echo has not been trivial, requiring the high resolution offered by the {\it Hubble Space Telescope (HST)}~or ground-based adaptive optics (AO).  Here we report detections of SN 2006gy using {\it HST}~and Keck AO at $\sim3000$ days post-explosion and consider the emission mechanism for the very late-time light curve.  While the optical light curve and optical spectral energy distribution are consistent with a continued scattered-light echo, a thermal echo is insufficient to power the $K'$-band emission by day 3000.  Instead, we present evidence for late-time infrared emission from dust that is radiatively heated by CSM interaction within an extremely dense dust shell, and we consider the implications on the CSM characteristics and progenitor system.
\end{abstract}


\section{Introduction}
\label{sec:intro}

\begin{table*}
\centering
\caption{{\it HST} Observations \label{tab1}}
\begin{tabular}{ l c c c c c c c c}
\hline
UT Date & Epoch & Program & PI & Instrument & Grating/Filter & Central $\lambda$ & Exp. & Magnitude\\
	       &(days)  &                  &     &                     &                          & (\AA)                            & (s)              & (Vega)\\
\hline
\multirow{2}{*}{20121216} & \multirow{2}{*}{2303} & \multirow{2}{*}{13029} & \multirow{2}{*}{Filippenko} & WFC3/UVIS & F625W & 6241 & 510 & 21.404 (0.013)\\
& & & & WFC3/UVIS & F814W & 8026 & 680 & 20.948 (0.014)\\
\hline
\multirow{3}{*}{20130302} & \multirow{3}{*}{2379} & \multirow{3}{*}{13025} & \multirow{3}{*}{Levan} & WFC3/UVIS & F275W & 2707 & 846 & $> 25.3$\\
& & & & WFC3/UVIS & F390W & 3922 & 932 & 22.882 (0.017)\\
& & & & WFC3/IR & F160W & 15369 & 206 & $> 14.7$ \\
\hline
\multirow{4}{*}{20141114} & \multirow{4}{*}{3001} & \multirow{4}{*}{13287} & \multirow{4}{*}{Fox} & STIS/CCD & G750L & 7751 & 5228 & --- \\
& & & & STIS/CCD & G430L & 4300 & 5848 & ---\\
& & & & STIS/MAMA & MIRNUV & 2305 & 697 & $>27.0$ \\
& & & & STIS/MAMA & MIRFUV & 1452 & 1957 & $>24.8$\\
\hline
\end{tabular}
\end{table*}

At the time of its discovery, supernova (SN) 2006gy was the most luminous SN ever observed \citep{quimby06,prieto06,ofek07,smith07}.  The Type IIn spectrum (see \citealt{filippenko97} for a review of SN spectral classification) indicated that interaction with dense circumstellar material (CSM) might play an important role in its energetics, but interpreting this event was not straightforward.  Aside from its peak luminosity, SN 2006gy has several other distinguishing characteristics.  For example, the SN took nearly 70 days to rise to peak, remained brighter than $-21$ mag for $\sim100$ days, and had a total radiated energy of $>10^{51}$~erg.  

A number of studies attempt to explain the early-time properties of both SN 2006gy and the new class of ``superluminous supernovae'' (SLSNe) that emerged following its discovery, focusing on either a standard core-collapse event within dense CSM or the radioactive decay from several solar masses of $^{56}$Ni generated in a pair-instability explosion \citep[see][and references therein]{gal-yam12}.  Late-time observations offer an opportunity to rule out the pair-instability model with energetic constraints.  Limited ground-based adaptive optics (AO) and {\it Hubble Space Telescope (HST)}~photometry at $\sim300$--800 days are significantly lower than predicted by the $>10$ \msolar~of $^{56}$Ni pair-instability model that would have been required to power the light curve at peak \citep{nomoto07, smith08gy}.  While the day $\sim$300 photometry is consistent with radioactive heating from a minimum of 2.5 \msolar~of $^{56}$Ni, subsequent optical and infrared (IR) photometry reveals an observed excess of emission over the expected radioactive decay rate, suggesting that the decline cannot be explained by $^{56}$Co alone \citep{miller10}.   Given the seeming lack of evidence for CSM interaction at earlier times, amongst other reasons, \citet{smith08gy} propose the most likely scenario is a thermal-IR echo from a massive ($\sim0.1$ \msolar) dust shell located $\sim1$ ly from the SN that was radiatively heated by the SN peak luminosity, but they cannot rule out a large $^{56}$Co luminosity that was reabsorbed by dust.  \citet{miller10} later eliminate a large $^{56}$Co mass with Keck AO observations that show a slower decline in the IR luminosity than expected from radioactive decay.  A relatively blue late-time optical colour from those epochs also suggests the presence of a scattered-light echo.  These results rule out various pair-instability SN models.


Thermal light echo models predict the evolution of the IR light curve \citep{dwek83}.  A scattered-light echo should have an optical spectrum reminiscent of the SN spectrum at peak (as opposed to late-time CSM interaction).  At only $\sim1.2$\arcsec\ from the bright galaxy nuclear bulge with a prominent dust lane, however, late-time ($>300$ day) data for SN 2006gy have been relatively difficult to obtain.  No spectrum of SN 2006gy has been published since day 237.  [\citet{smith08gy} published a two-dimensional (2D) ground-based Keck spectrum of SN 2006gy on day 364, but a reliable 1D extraction was too difficult given the overwhelming galactic nucleus.]


Here we present ongoing observations of SN 2006gy through day 3024 to test for the presence and expected evolution of both a scattered and thermal light echo.  Data include day 3024 Keck/NIRC2-AO-LGS $K'$-band photometry, a day 3001 {\it HST}/STIS spectrum of SN 2006gy, and {\it HST}/WFC3 and STIS visible and ultraviolet (UV) photometry.  Section \ref{sec:obs} presents the observations.  Sections \ref{sec:optical} and \ref{sec:echo} analyse the data in the context of a light echo, while Section \ref{sec:csm} considers the case of CSM interaction.  Section \ref{sec:discussion} presents the implications of the results and summarises our conclusions.  

Throughout this paper we assume that the distance to NGC 1260 (the host galaxy of SN 2006gy) is 73.1 Mpc (redshift $z = 0.018$), and following \citet{smith07} we adopt $E(B - V ) = 0.54$ mag as the reddening toward SN 2006gy within the host galaxy, while Galactic extinction accounts for $E(B - V ) = 0.18$ mag, leading to a total colour excess toward SN 2006gy of $E(B - V ) = 0.72$ mag. Unless otherwise noted, all spectral energy distributions (SEDs) and spectra have been corrected for this colour excess assuming $R_V = A_V / E(B - V ) = 3.1$ using the reddening law of \citet{cardelli89}.

\section{Observations}
\label{sec:obs}

 \begin{figure*}
\centering
\includegraphics[width=2.3in]{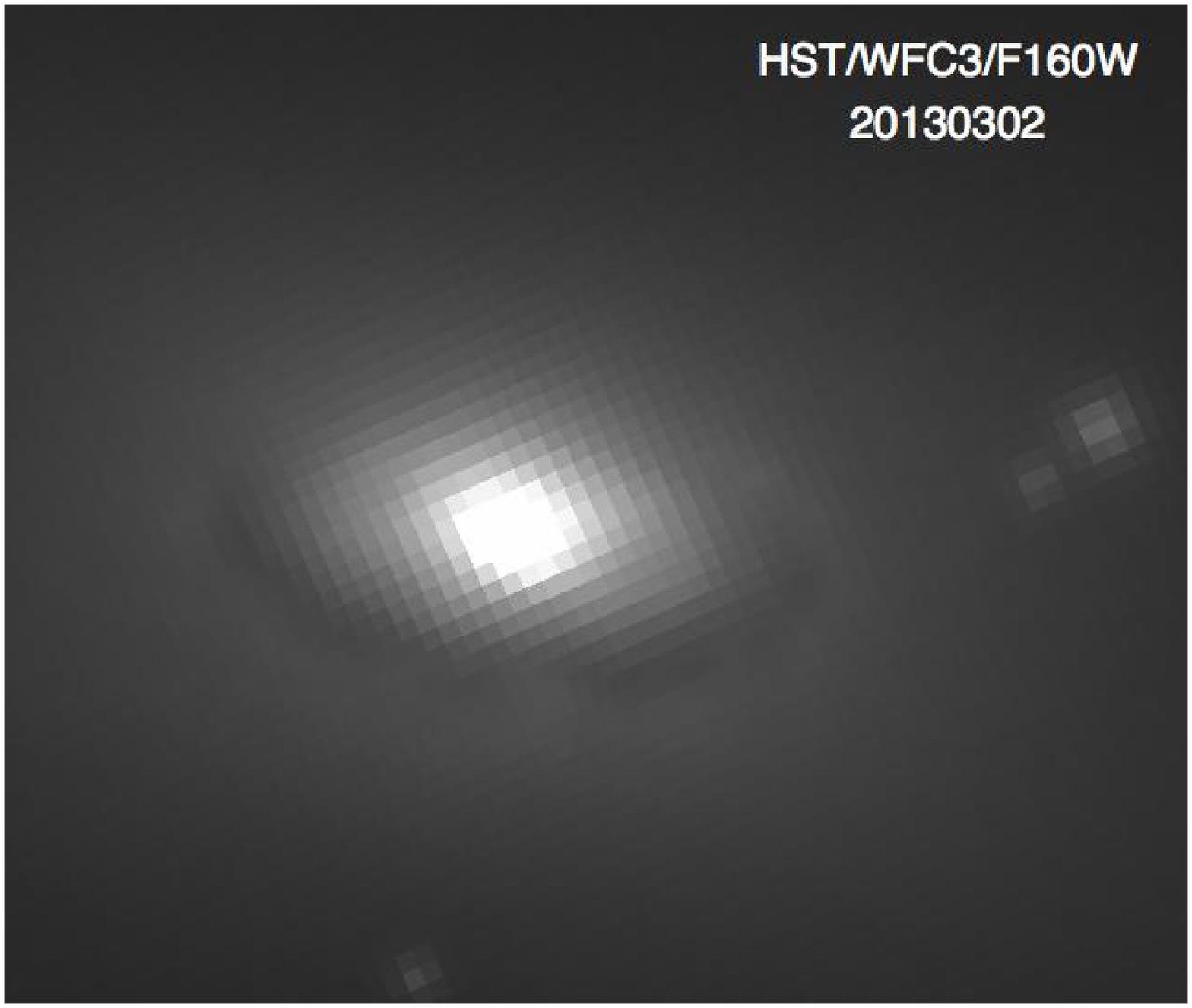}
\includegraphics[width=2.3in]{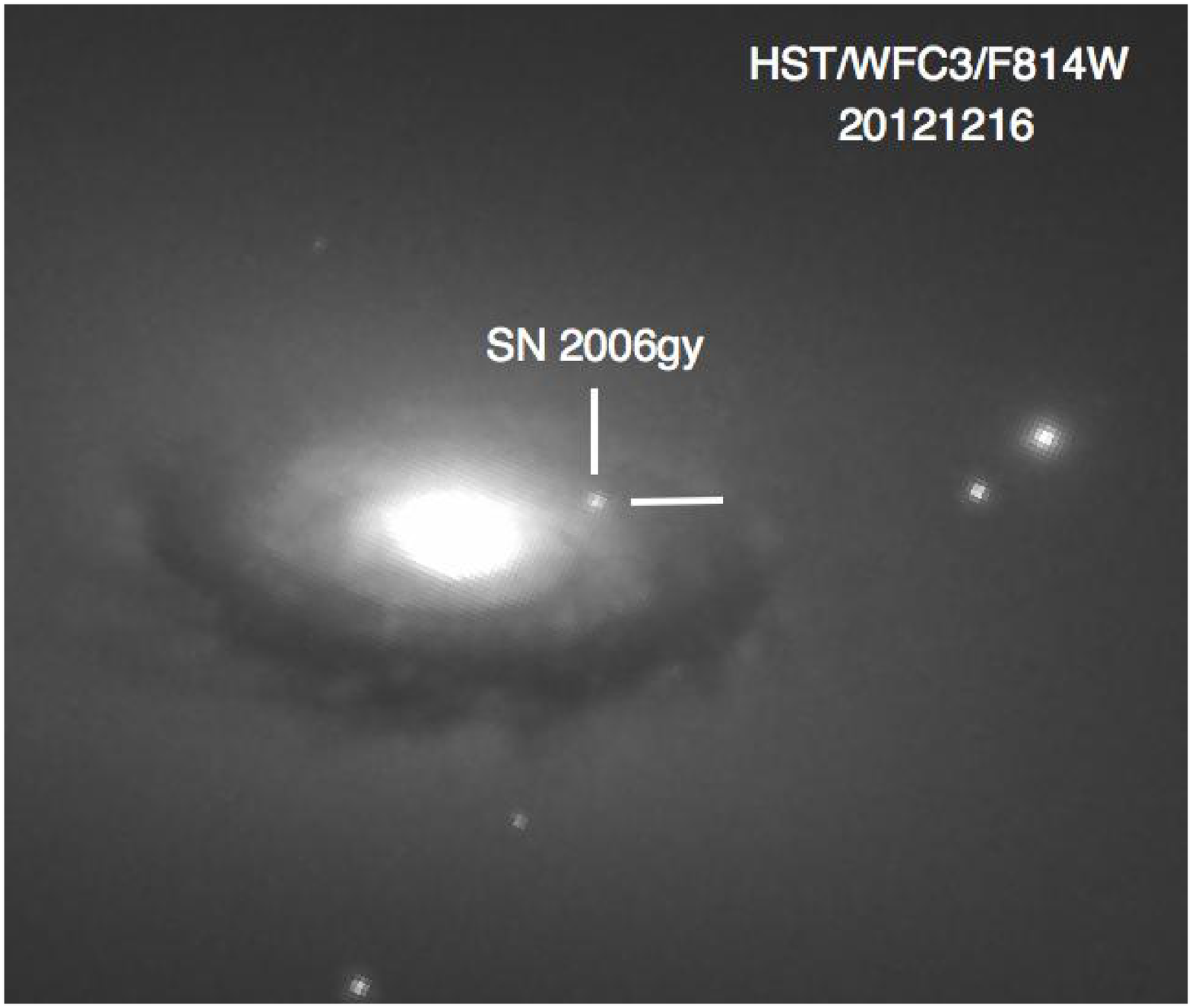}
\includegraphics[width=2.3in]{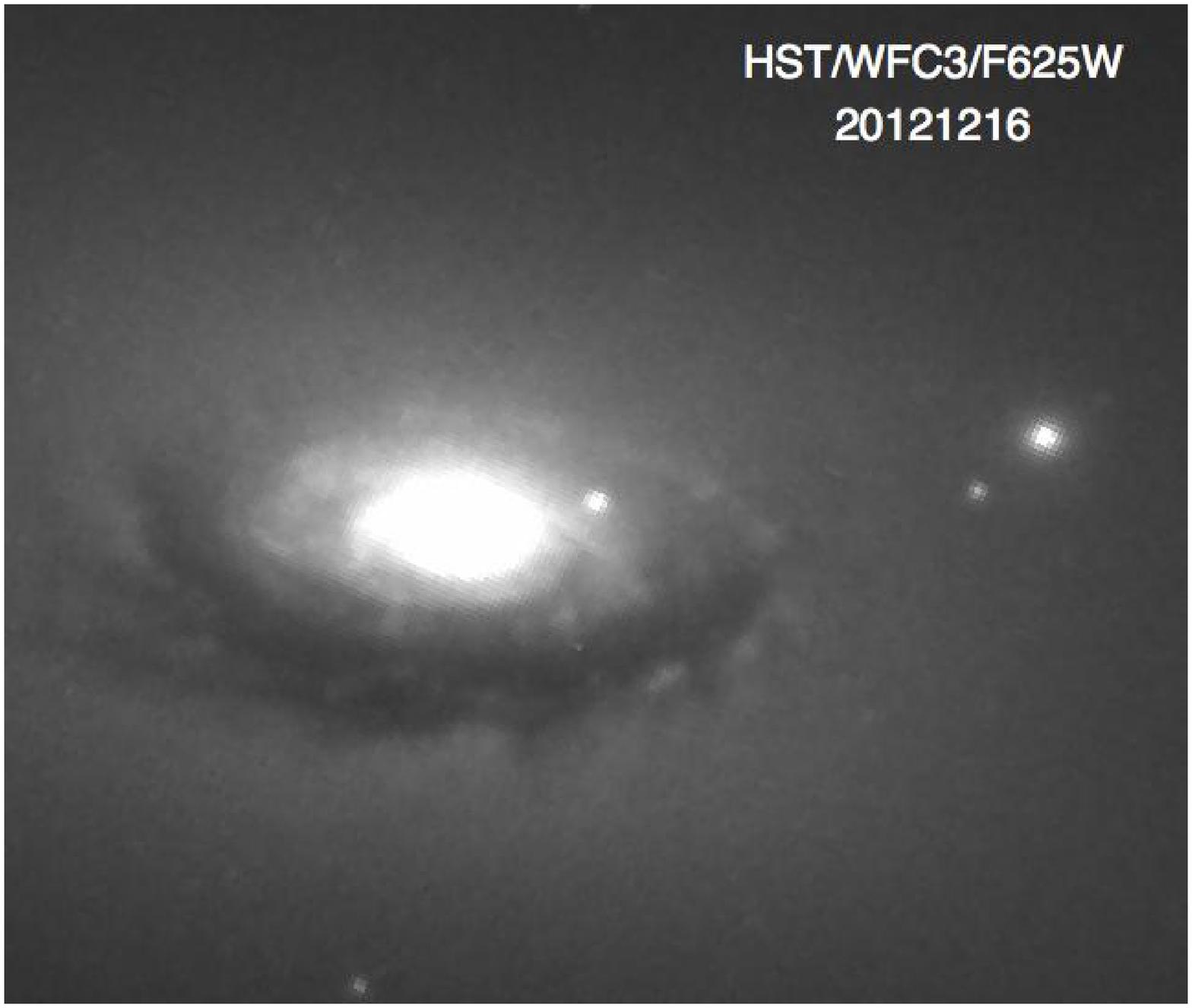}
\includegraphics[width=2.3in]{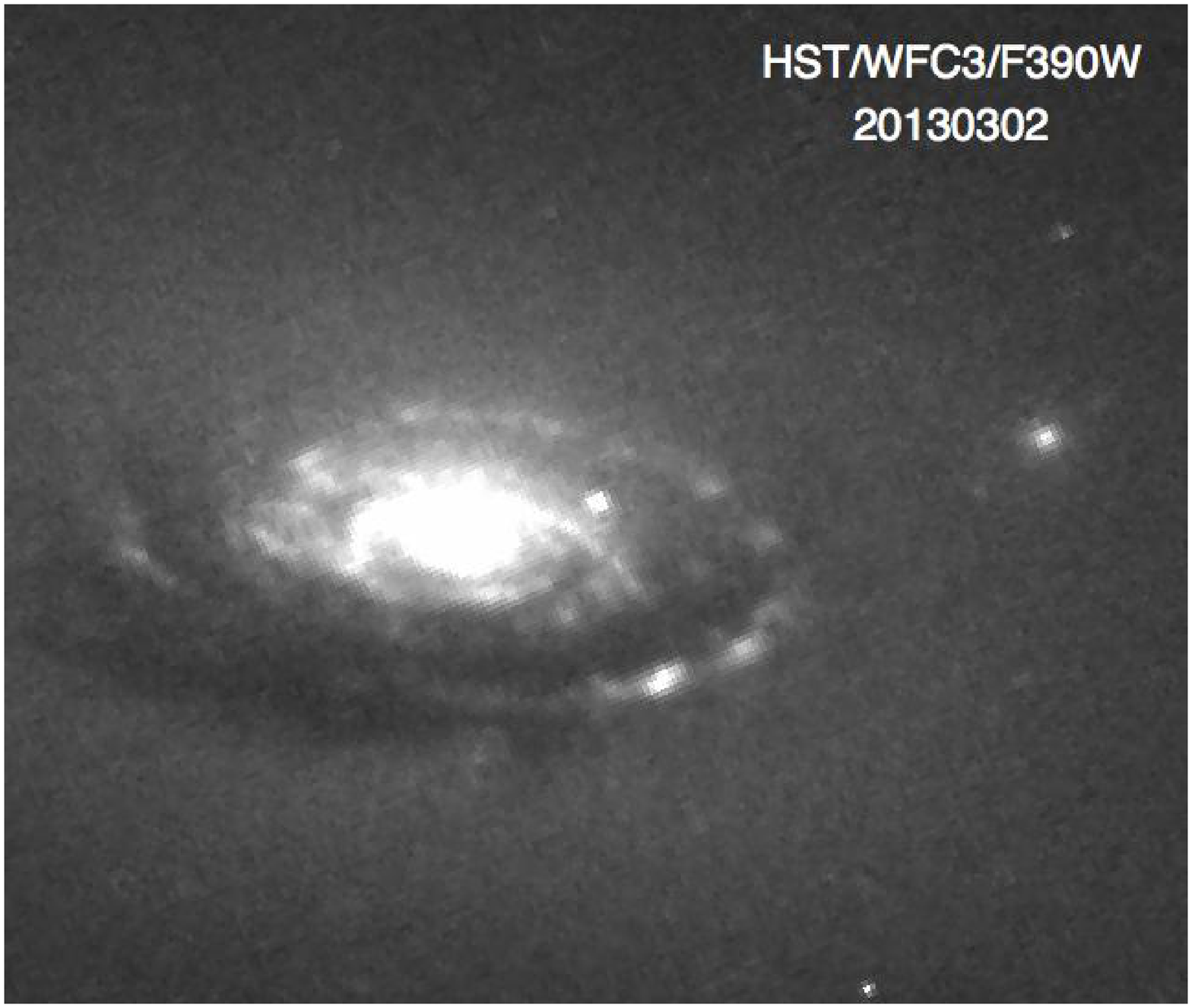}\\
\includegraphics[width=2.3in]{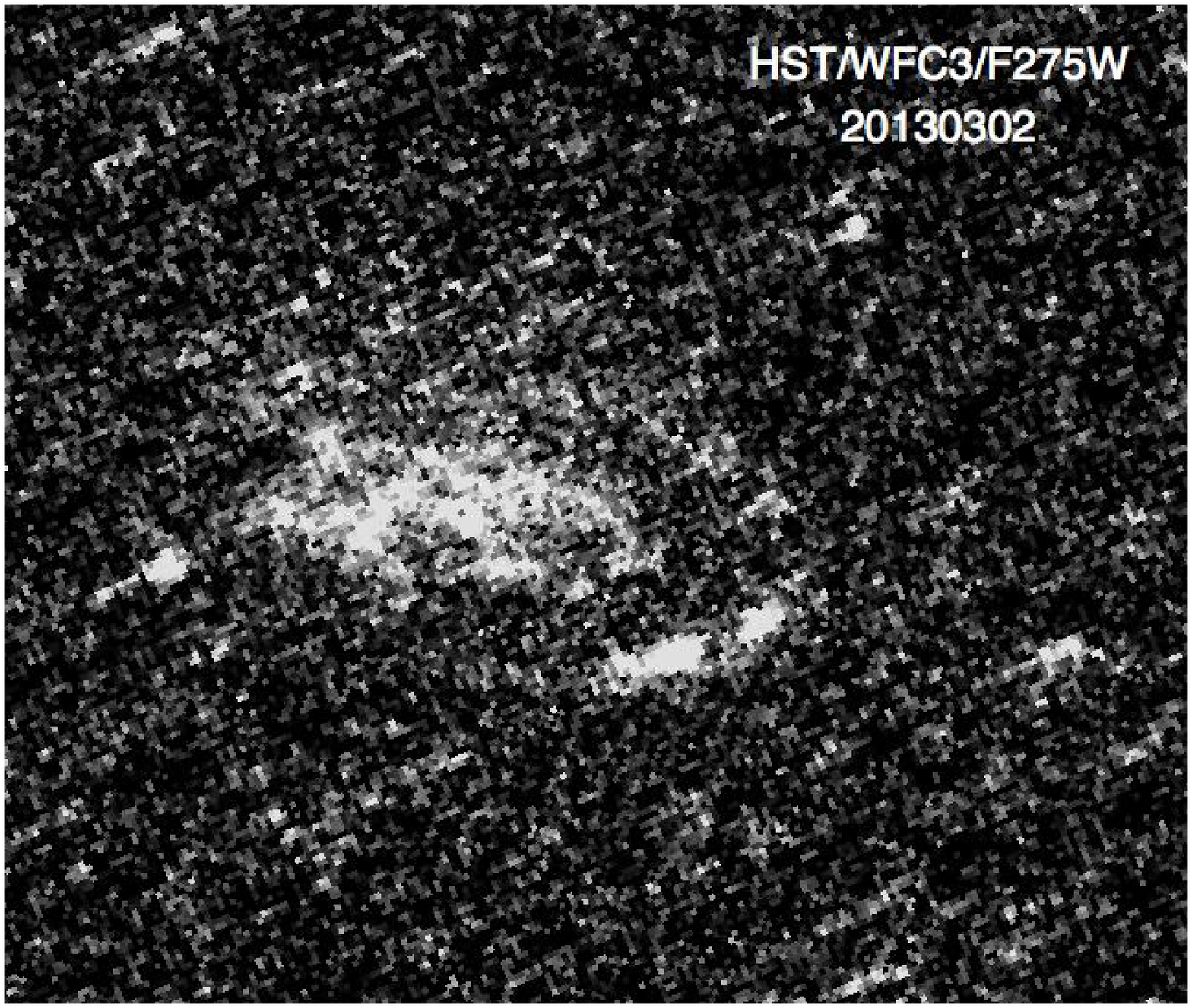}
\includegraphics[width=2.3in]{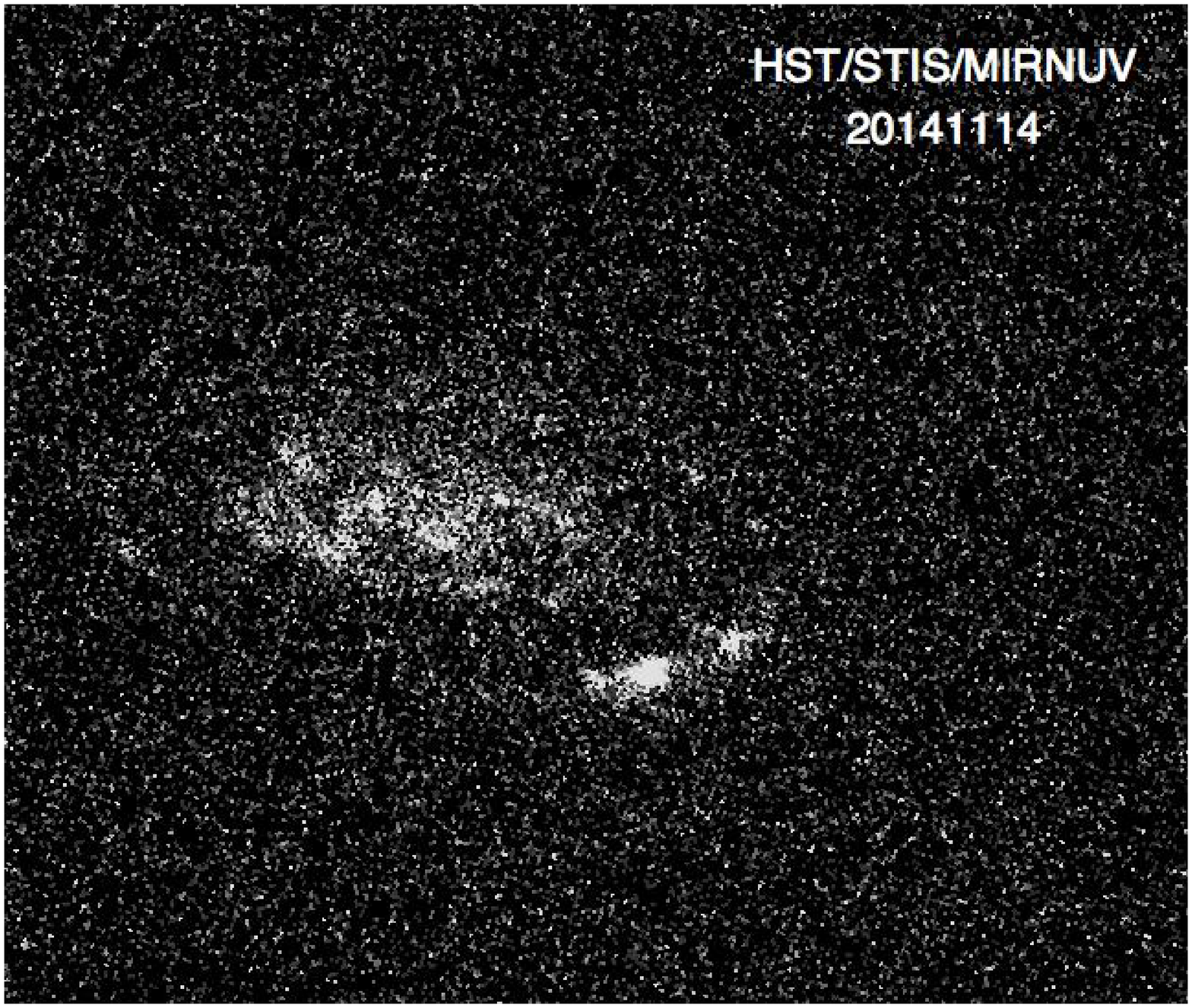}
\includegraphics[width=2.3in]{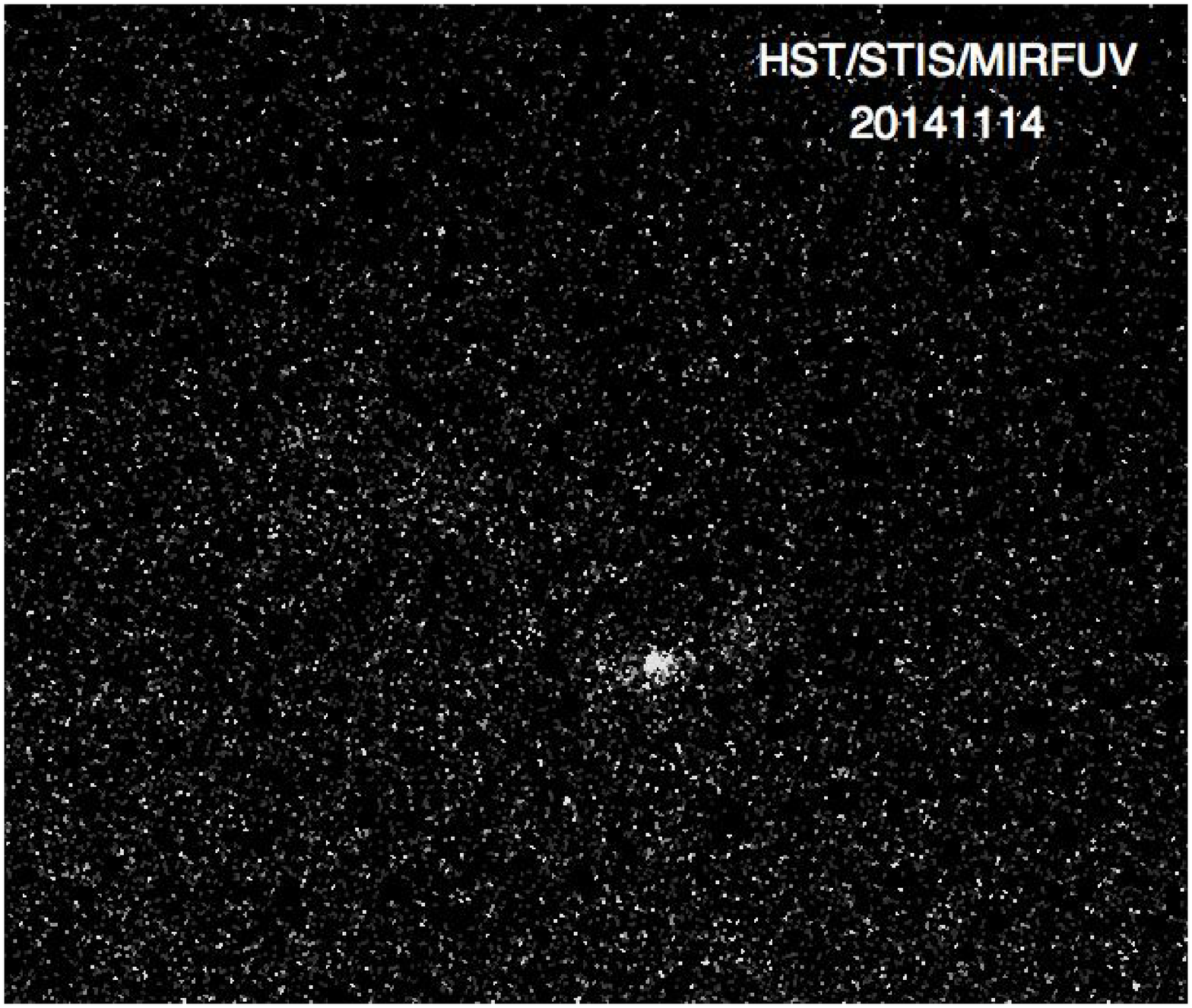}
\caption{{\it HST}~imaging of SN 2006gy.  Details of the instrument, filter, and observation date are labeled in each image and listed in Table \ref{tab1}.}
\label{fig_imaging}
\end{figure*}

\subsection{Hubble Space Telescope}
\label{sec:hst}

Table \ref{tab1} lists details regarding new data on SN 2006gy observed by {\it HST} as part of programs GO-13287 (PI Fox), GO-13025 (PI Levan), and GO-13029 (PI Filippenko).

\subsubsection{WFC3 Photometry}

SN 2006gy was observed with the {\it HST}/WFC3 UVIS and IR channels, summarised in Table \ref{tab1} and displayed in Figure \ref{fig_imaging}.  Photometry was extracted from the individual WFC3 ``\_flt.fits'' images in all bands using Dolphot v2.0 \citep{dolphin00}.  The input parameters are those recommended by the Dolphot WFC3 Users' Manual.  Aperture corrections were applied.  The region in the F160W filter is too confused for accurate photometry.  Instead, the upper limit is based on unreal sources that Dolphot thinks it has detected in the main body of the galaxy, near the nucleus and the dust lane.  Table \ref{tab1} lists the resulting magnitudes in the WFC3 flight system (Vegamag).


\subsubsection{STIS Photometry}

SN 2006gy was observed with the {\it HST}/STIS NUV and FUV MAMA channels, summarised in Table \ref{tab1} and displayed in Figure \ref{fig_imaging}.  No source is detected at the position of the SN (a bright knot in one the galactic arms is all that is visible).  We calculate upper limits using standard photometry techniques in {\tt IRAF}'s APPHOT package and convert to magnitudes using procedures outlined by the STIS Data Handbook Section 5.3.

\subsubsection{STIS Spectroscopy}

\begin{figure*}
\centering
\includegraphics[width=3.6in]{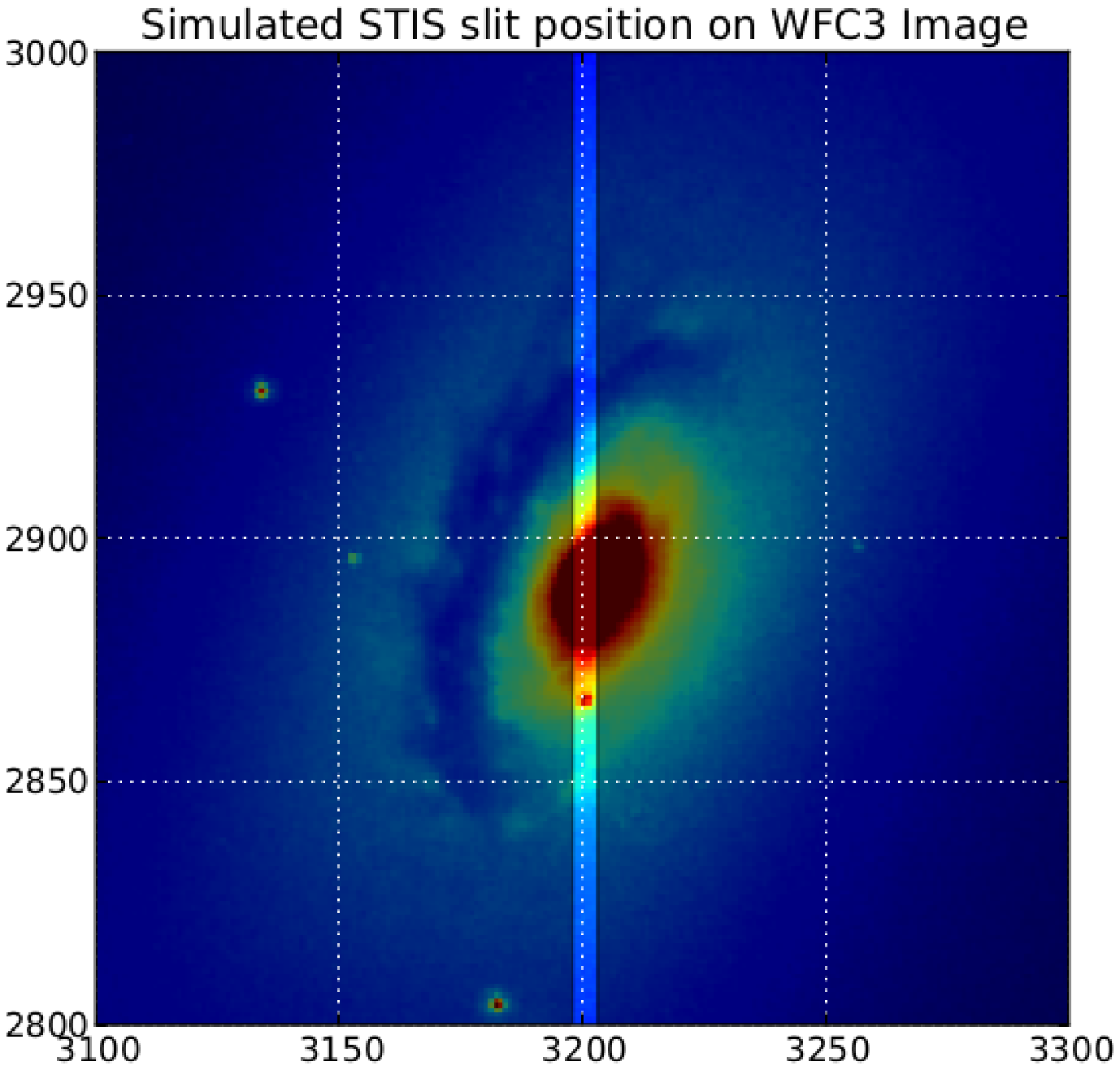}
\hspace{-20pt}
\includegraphics[width=3.5in]{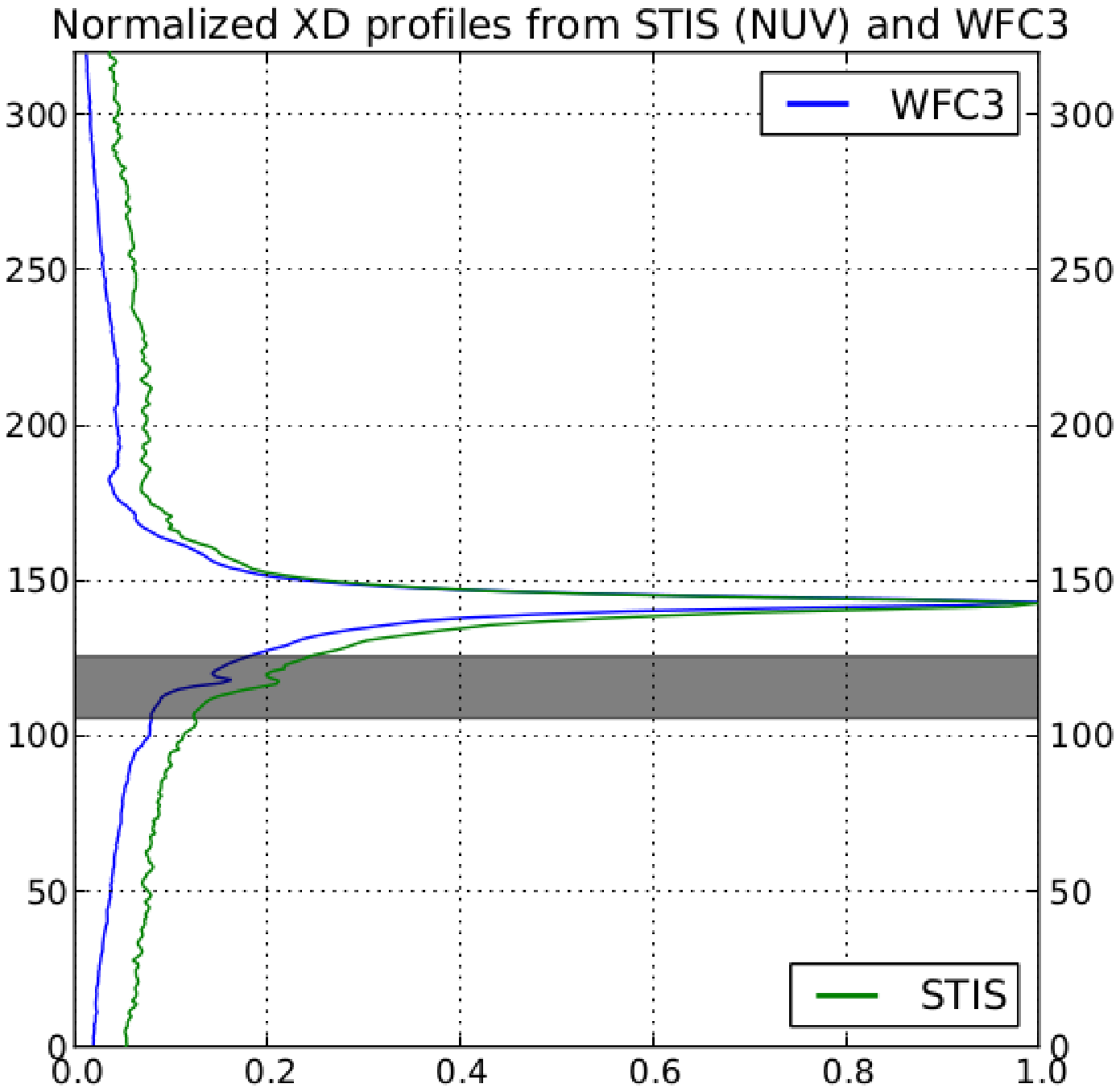}
\caption{Slit position orientation confirmation.  {\bf (left)} The {\it HST}/WFC3/F814W image with the planned slit orientation highlighted.  {\bf (right)} A simulated cross-dispersion profile (blue) created by summing rows within the slit on the left-hand side.  The green line represents the actual STIS cross-dispersion profile, calculated by median-stacking the columns in the pipeline-reduced STIS spectrum (ocdd04010\_crj.fits).  The similarity of these two profiles indicates the SN, highlighted by the grey bar just south of the galaxy nucleus, is centred in the slit.}
\label{fig_finder}
\end{figure*}

 Table \ref{tab1} summarises the optical spectra of SN 2006gy obtained with {\it HST}/STIS.  Given that the SN magnitude was unknown at the time of the observations, the SN was centred in the $52 \times 0.2$\arcsec\ slit by offsetting from a nearby star.  The location of the supernova in the 2D image is identified by matching a simulated cross-dispersion (XD) profile from WFC3/F814 observations (day 2303) with the actual XD profile of the STIS/G750L spectrogram.  We choose these two observations for comparison since they are most closely matched in wavelength coverage and include the prominent H$\alpha$~line.  

Although we specified the slit position on the sky in the original Phase 2 files, we consider a range of actual slit positions.  For each position, the WFC3 XD profile is simulated by summing the WFC3 pipeline calibrated image (\_drz.fits) in the dispersion direction over $\sim5$ pixels, which corresponds to the STIS slit width on the WFC3 detector. The STIS XD profile is created by collapsing the 2D spectrogram in position space and rebinning to the WFC3 plate scale.  The slit position is then constrained by matching the WFC3 and STIS XD profiles (see Figure \ref{fig_finder}).

\begin{figure*}
\centering
\includegraphics[width=6.3in]{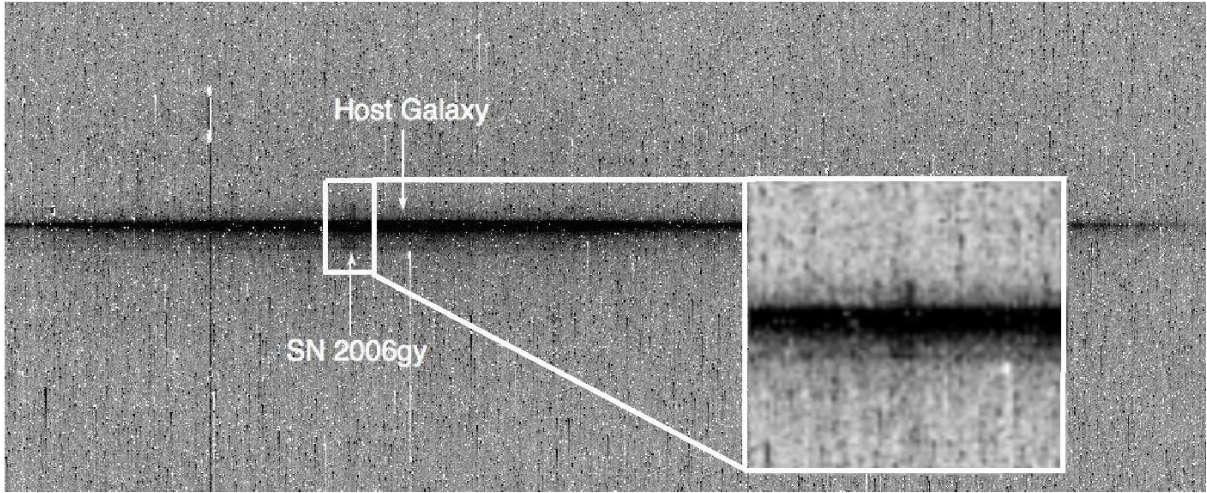}
\caption{The 2D spectrum of SN 2006gy obtained with the G750L grating (ocdd04010\_crj.fits).  The SN is highlighted by the faint trace just below the bright galaxy nucleus.  The H$\alpha$~line, highlighted by the inset, is the most obvious feature.}
\label{fig_2d}
\end{figure*}

\begin{figure*}
\centering
\includegraphics[width=7.7in]{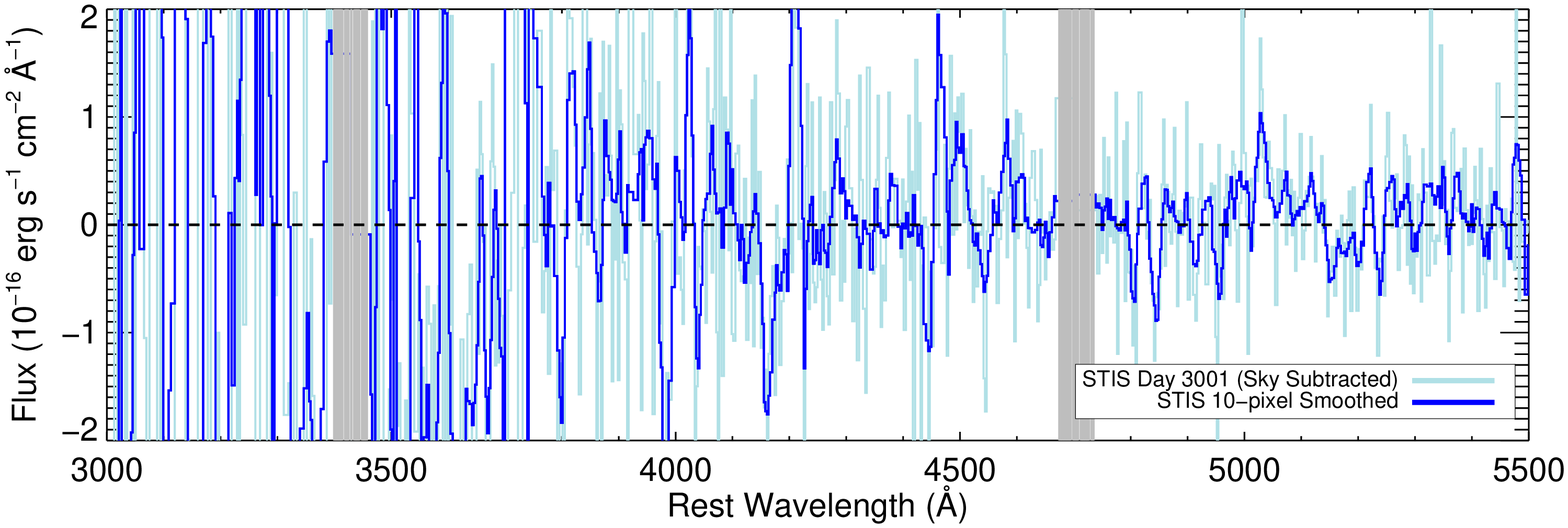}\\
\includegraphics[width=7.7in]{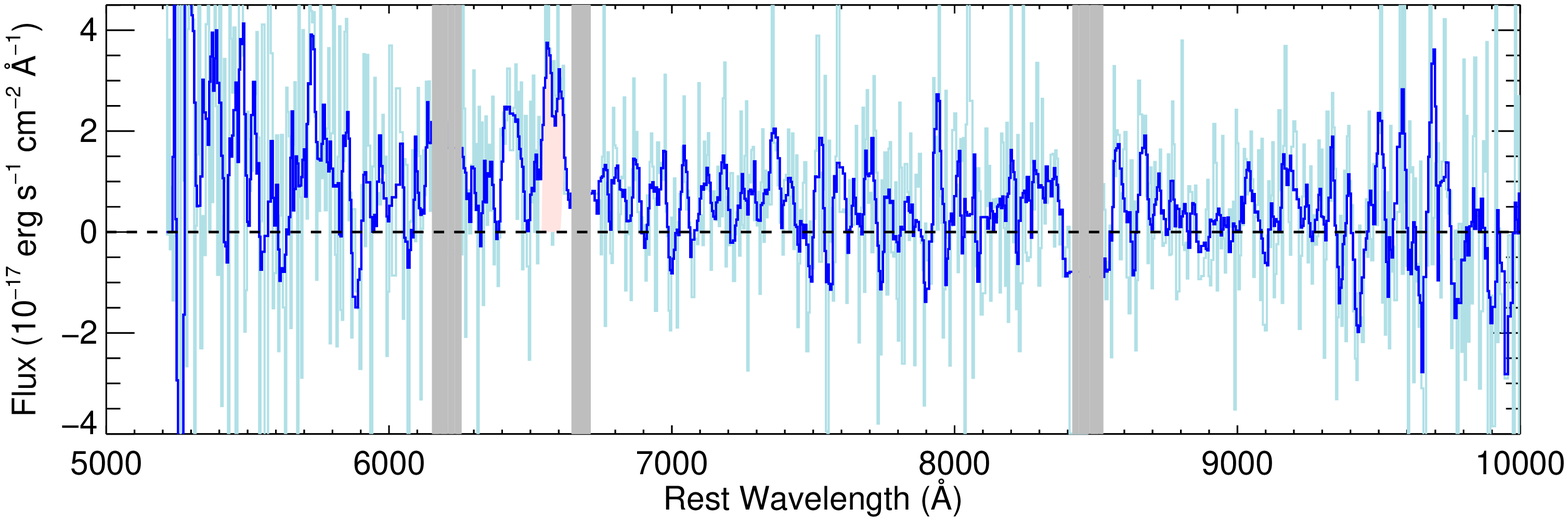}
\caption{Background-subtracted {\it HST}/STIS spectrum of SN 2006gy on day 3001 post-explosion, corrected for extinction assuming a reddening of $E(B - V) = 0.72$ mag.  The light blue plots the unsmoothed spectrum, while the dark blue plots the spectrum that has been 10-pixel boxcar smoothed.  Grey bars signify pixels flagged for having bad data quality.  The spectrum is relatively featureless, flat, and noisy.  We detect H$\alpha$~(shaded red), but no other obvious lines can be identified.}
\label{fig_stis_spectrum} 
\end{figure*}

The 1D spectrum for each observation is extracted using the CALSTIS custom extraction software stistools.x1d\footnote{http://ssb.stsci.edu/doc/stsci\_python\_x/stistools.doc/html/x1d.html?highlight=x1d\#module-stistools.x1d}.  The default extraction parameters for STIS are defined for an isolated point source. For both the G430L and G750L the default extraction box width is 7 pixels and the background extraction box width is 5 pixels.  SN 2006gy, however, is quite faint and located just a few pixels in position space below the galaxy (0.05\arcsec pixel$^{-1}$; Figure \ref{fig_2d}).  We therefore reduce the extraction box width to 3 pixels to optimise the signal-to-noise ratio (S/N).  After locating the SN position along the slit above, the optimal extraction pixel position was determined by shifting the extraction box along the 2D spectrogram in position space and optimising the H$\alpha$~S/N.  The three extractions above and below this SN position were median combined to estimate the background.  All of the STIS spectra are combined to produce a single spectrum using the splice tool in STSDAS\footnote{http://stsdas.stsci.edu/cgi-bin/gethelp.cgi?splice}.  We ignore the edge columns by setting their data-quality flags to 4. Figure \ref{fig_stis_spectrum} plots the final, background-subtracted spectrum.

\subsection{Keck/NIRC2-AO-LGS}
\label{sec:keck}

SN 2006gy was observed with Keck/NIRC2-AO-LGS \citep{wizinowich06} on 2014 Dec. 7 UT using the wide-field camera and $K'$~filter.  The complete set of observations include two repeats of the ``bxy9'' pattern with 1\arcsec~dithers, where each frame consisted of three coadded 8\,s exposures (i.e., 24\,s frame).  To reduce the data, we subtracted a median bias frame, applied flat-field corrections to each exposure, and corrected for astrometric distortion\footnote{https://www2.keck.hawaii.edu/inst/nirc2/dewarp.html}.  A bright star present in all the images was used to align and, ultimately, coadd the dithered images.

The resulting data were analysed using standard IR analysis techniques utilising {\tt SExtractor}\footnote{SExtractor can be accessed from http://www.astromatic.net/software.}.  Given the steep gradient of the underlying galaxy, we chose to subtract this contribution by using GALFIT \citep{peng02,peng10} to model NGC 1260 with a radial Sersic (\citeyear{sersic63}) profile.  We use a bright, isolated star as a model for the point-spread function. We allowed the exponent $n$, half-light radius, axis ratio, position angle, galaxy position, and sky background to vary.  Removing the principal source of variation in the background yields a more robust local background subtraction.  Calibration was performed using field stars with reported fluxes in 2MASS \citep{skrutskie06}.  Table \ref{tab_nirc2_phot} lists the new photometry and that reported by \citet{miller10}.  Figure \ref{fig_06gy_lc} plots this photometry along with the WFC3 photometry presented in Table \ref{tab1}.

\subsection{Keck Optical Spectroscopy}

For a comparative analysis in Section \ref{sec:csm}, we also present previously unpublished optical spectra of SNe IIn 2005ip and 2010jl at similar late-time epochs, summarised in Table \ref{tab_opt_spectra}.  The spectra were obtained with the dual-arm Low Resolution Imaging Spectrometer \citep[LRIS;][]{oke95} mounted on the 10-m Keck~I telescope with the slit aligned along the parallactic angle to minimise differential light losses \citep{filippenko82}.  The spectra were reduced using standard techniques \citep[e.g.,][]{foley03,silverman12bsnip1}. Routine CCD processing and spectrum extraction were completed with {\tt IRAF}\footnote{IRAF: the Image Reduction and Analysis Facility is distributed by the National Optical Astronomy Observatory, which is operated by the Association of Universities for Research in Astronomy (AURA), Inc., under cooperative agreement with the US National Science Foundation (NSF).}, and the data were extracted with the optimal algorithm of \citet{horne86}. We obtained the wavelength scale from low-order polynomial fits to calibration-lamp spectra. Small wavelength shifts were then applied to the data after cross-correlating a template-sky spectrum to an extracted night-sky spectrum. Using our own IDL routines, we fit a spectrophotometric standard-star spectrum to the data in order to flux calibrate the SN and to remove telluric absorption lines \citep{wade88,matheson00}.

\begin{table}
\centering
\caption{Keck AO Observations of SN 2006gy \label{tab_nirc2_phot}}
\begin{tabular}{c c c c}
\hline
Date & Epoch & Filter & Mag\\
(UT) & (days)  &           &  (Vega)\\
\hline
2007 Sep. 29  & 398  & $K'$  & 14.91 $\pm$ 0.17 \\
2007 Dec. 2    &  461  & $H$ & 16.8 $\pm$ 0.3 \\
2007 Dec. 2    & 461  & $K'$ & 15.02 $\pm$ 0.17 \\
2008 Aug. 25  & 723  & $K'$  & 15.59 $\pm$ 0.21\\
2014 Dec. 07 & 3024 & $K'$ & 18.10 $\pm$ 0.17\\
\hline
\end{tabular}
\end{table}

\begin{table}
\centering
\small
\caption{Summary of Keck/LRIS Optical Spectra \label{tab_opt_spectra}}
\begin{tabular}{ l c c c c}
\hline
SN & JD $-$ & Epoch & Res. & Exp.\\
      & 2,450,000 & (days) & (\AA) & (s) \\
\hline
2005ip & 6778 & 3024 & $\sim6$ & 1200 \\   
2010jl  & 6778  & 1290 & $\sim6$ & 1200 \\ 
\hline
\end{tabular}
\end{table}

\section{A Scattered Optical Light Echo?}
\label{sec:optical}

The observed late-time optical emission from SN 2006gy has been attributed previously to a scattered-light echo \citep{smith08gy,miller10}, which is a product of scattered light from the SN light curve emerging from a paraboloid of revolution with the supernova as its focus and its axis along the line of sight \citep[e.g.,][]{bode80,dwek83,chevalier86}.  We consider this scenario in the context of the new data presented in this article.

Figure \ref{fig_opt_sed} plots the optical SED of SN 2006gy on day 2379, which corresponds to the single epoch of F390W observations.  Although data through the other filters were not obtained at this particular epoch, we extrapolate the F625W and F814W photometry shown in Figure \ref{fig_06gy_lc}.  This figure goes on to compare the photometry to the expected scattered echo spectrum, which is the cumulative scattering of the whole light curve.  To simulate this spectrum, we construct a synthetic integrated spectrum by mean-combining the individual spectra from day 36 (pre-peak), day 71 (peak), day 122, and day 177 (spaced roughly every 40 days; \citealt{smith10}).  We then assume a $\lambda^{-0.95}$~wavelength dependence for the scattering, which can be considered typical \citep[e.g.,][]{miller10}.  Since the flux of the scattered-light echo depends on the specific arrangement of the dust, the precise scale factor is unknown.  Instead, Figure \ref{fig_opt_sed} scales the spectra to the F814W photometry.  Overall, the synthetic photometry for a scattered-light echo with a $\sim\lambda^{-0.95}$~wavelength dependence is consistent with the observed SED on day 2379.

\begin{figure}
\centering
\includegraphics[width=3in]{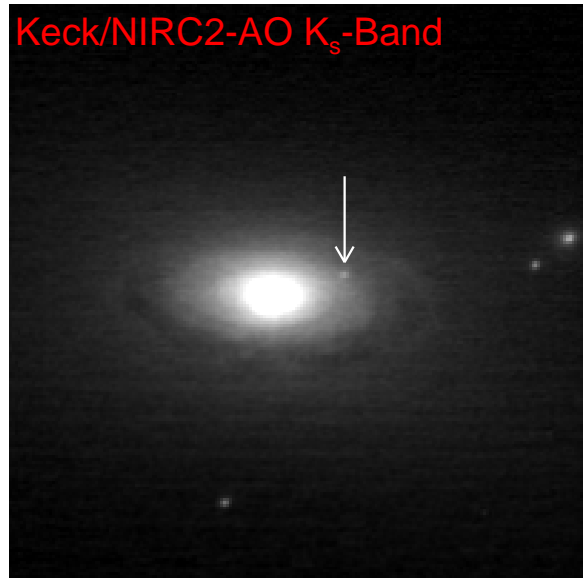}
\caption{Keck/NIRC2/LGS-AO image of SN 2006gy obtained on 2014 Dec. 7 UT.}
\label{fig_nirc2_img} 
\end{figure}

\begin{figure*}
\centering
\hspace{-0.5in}
\includegraphics[width=7.7in]{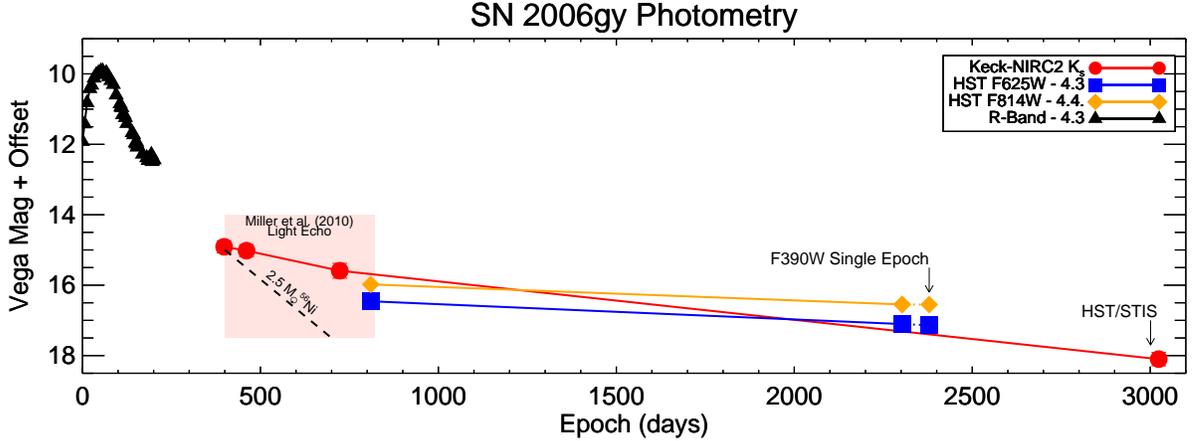}
\caption{Light curve of SN 2006gy through day 3024, including data from \citet{smith08gy} and \citet{miller10}.  Dotted lines for the F625W and F814W filters beyond day $\sim 2300$ illustrate the extrapolations used to calculate the photometry plotted in Figure \ref{fig_opt_sed}, corresponding to the single epoch of F390W photometry on day 2379.  Offsets are applied only for plot clarity.}
\label{fig_06gy_lc} 
\end{figure*}

\begin{figure*}
\centering
\hspace{-0.5in}
\includegraphics[width=7.7in]{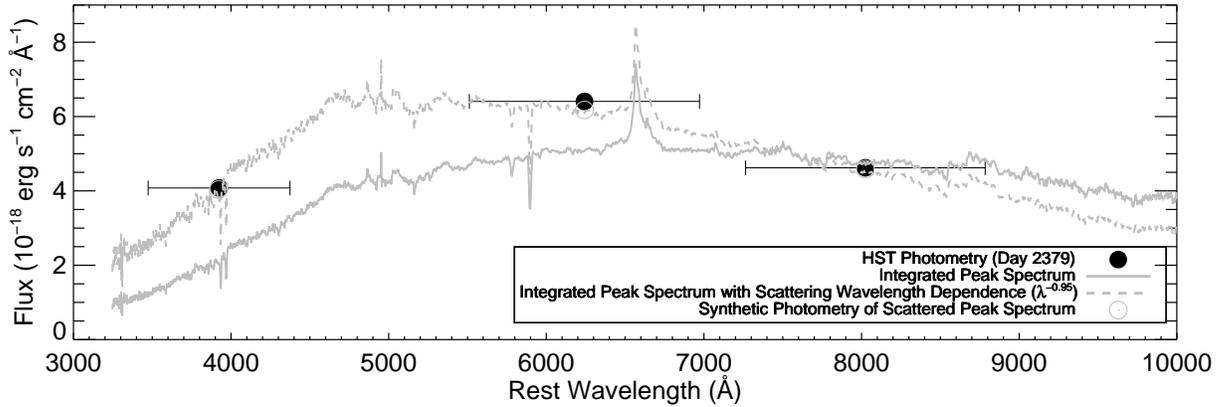}
\caption{The optical SED of SN 2006gy on day 2379, which corresponds to the single epoch of F390W observations.  Overplotted in grey (solid) is the synthetic integrated spectrum of SN 2006gy around peak light, constructed by mean-combining the day 36 (pre-peak), day 71 (peak), day 122, and day 177 data.  A $\lambda^{-0.95}$~wavelength dependence is assumed for the scattered-light peak spectrum (dashed grey).  Both spectra are scaled to the F814W photometry.  The synthetic photometry of the scattered spectrum is consistent with the observed SED.}
\label{fig_opt_sed} 
\end{figure*}

\section{A Thermal-IR Light Echo?}
\label{sec:echo}

\subsection{The Peak of the Thermal SED}
\label{sec:peak}

\begin{figure*}
\centering
\includegraphics[width=3.3in]{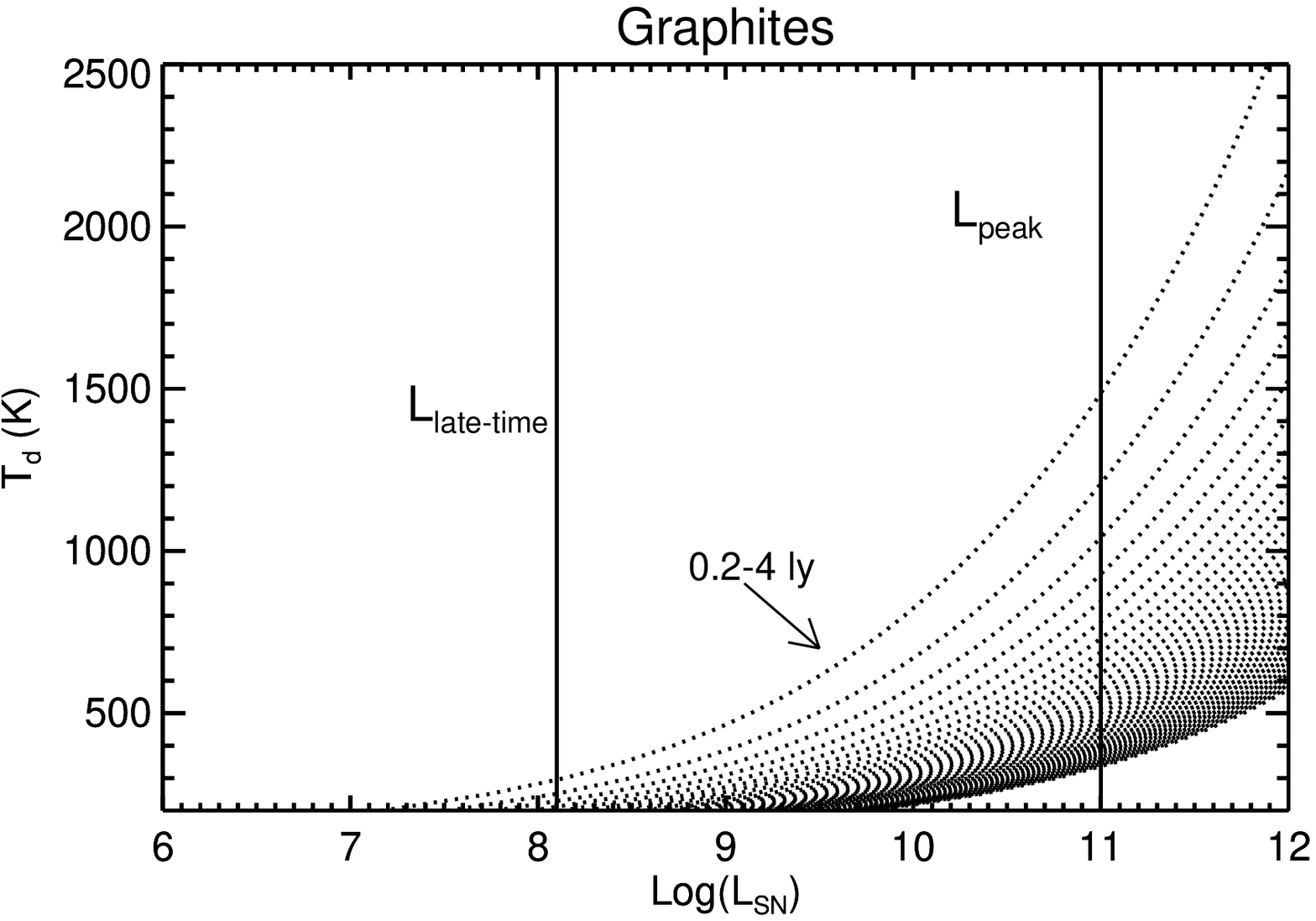}
\includegraphics[width=3.3in]{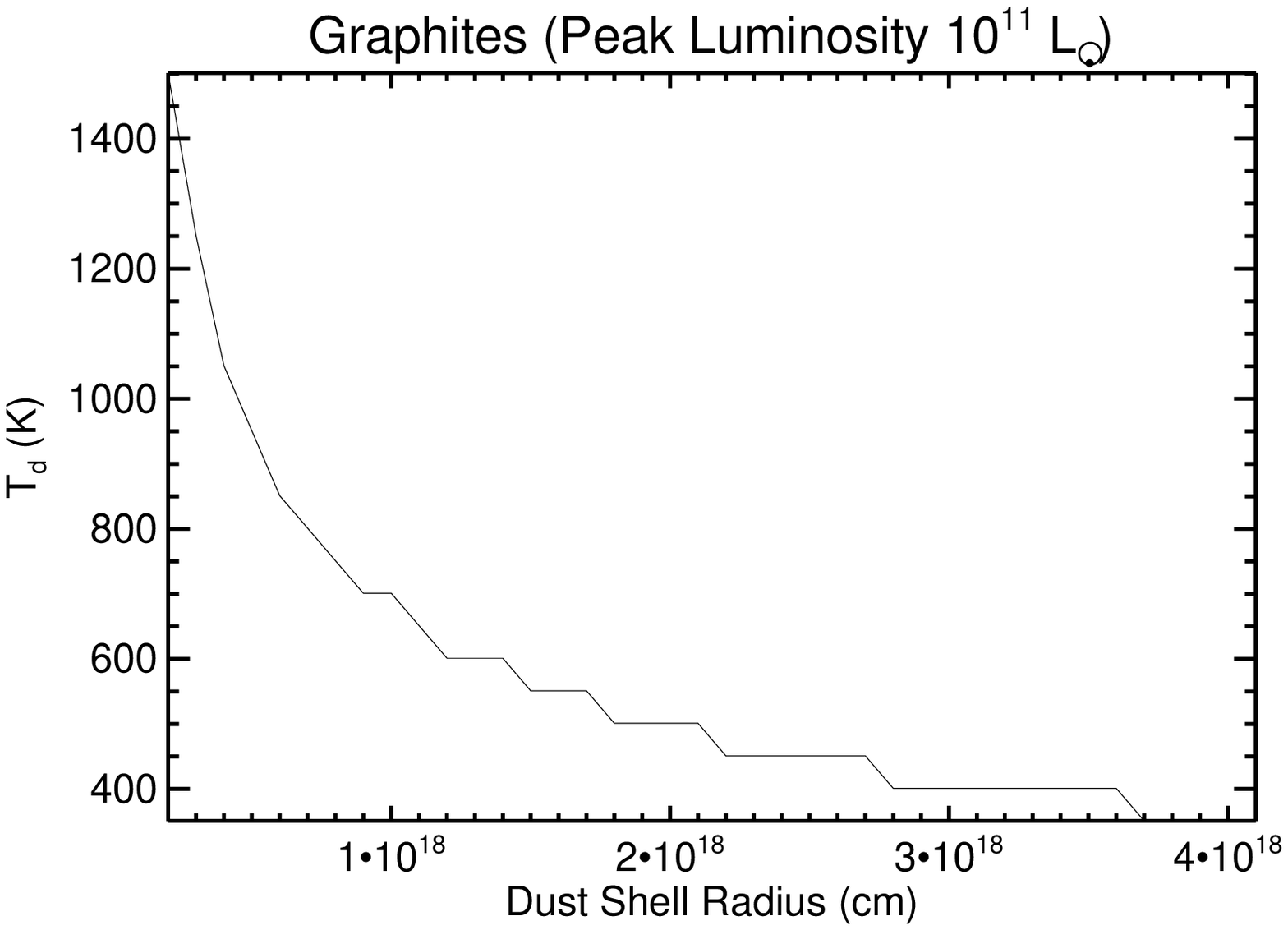}
\caption{Relationship between dust mass, dust temperature, and distance from the heating source.  {\bf (left)} The dust temperature as a function of the central energy source (i.e., SN peak luminosity) for 0.1 \micron~graphite dust grains at distances ranging from 0.2 to 4~ly (dotted lines).  Overplotted as a reference are the luminosities corresponding to both the peak and late-time plateau  of SN 2006gy.  {\bf (right)} The dust temperature as a function of distance for the specific case of the observed peak luminosity of SN 2006gy ($L_{\rm peak} \approx 10^{11}$~\lsolar).}
\label{fig_echotemp} 
\end{figure*}

Little late-time IR colour information exists for SN 2006gy given that the observations require high-resolution ground-based AO or {\it HST} photometry.  Prior to this article, just 1 $H$- and 4 $K'$-band observations existed at $>1$ yr post-explosion.  The equilibrium dust temperature and, thereby, the peak of the corresponding SED remain relatively unconstrained.  Both \citet{smith08gy} and \citet{miller10} place a {\it lower limit} on the dust IR luminosity by assuming that (1) all of the near-IR luminosity can be attributed to a thermal light echo, and (2) the near-IR luminosity peaks in the $K'$ band given the very red $H-K'$~colour observed on day $\sim400$.  Here we present a quantitative analysis of the $K'$-band contribution to the total IR luminosity in the thermal-IR echo scenario.

First, we calculate the equilibrium temperature of dust over a range of distances from the SN following dust heating models outlined by \citet{fox10}.  Assuming 0.1~\micron~graphite dust grains, Figure \ref{fig_echotemp}(a) plots the dust temperature as a function of the central energy source (i.e., SN peak luminosity) for dust at distances ranging from 0.2 to 4~ly.  For any given luminosity, the dust is heated to expectedly lower temperatures at larger distances.  Figure \ref{fig_echotemp}(b) plots the dust temperature as a function of distance for the specific case of the observed peak luminosity of SN 2006gy ($L_{\rm peak} \approx 10^{11}$~\lsolar).  

Since the dust temperature can be written as a function of radius from the SN, we can also plot the fraction of the total IR luminosity emitted in $K'$ as a function of radius (Figure \ref{fig_frack}).  At larger distances, where the dust temperature drops and the SED peak shifts to longer wavelengths, the fraction of the total IR flux emitted in $K'$ decreases.  

For the case of an IR echo, the quantitative relationship between observation epoch and the emitting dust shell radius is not straightforward because the paraboloid intersects small fractions of many thin shells at any given instant (see \citealt{dwek83}).  While modeling the integrated flux from the many thin shells is beyond the scope of this paper, the hottest dust at any epoch, $t$, is located at a radius $R=ct/2$, where $c$~is the speed of light.  This radius therefore sets the upper limit to the fractional $K'$-band emission (all other contributing shells in the paraboloid have lower temperatures).  By writing the observation epoch as a function of the hottest dust shell radius, $t=2R/c$, Figure \ref{fig_frack} also plots the maximum fraction of the total IR luminosity emitted in $K'$ as a function of time post-explosion.  The fraction of $K'$-band flux at a given epoch is independent of the CSM density or geometry.  At early times, the $K'$ flux represents only $\sim15$\%~of the total IR flux, and by late times, this fraction drops to $\lesssim$1\%.

\subsection{Other Potential $K'$-Band Flux Sources}
\label{sec_other}

\begin{figure}
\centering
\includegraphics[width=3.5in]{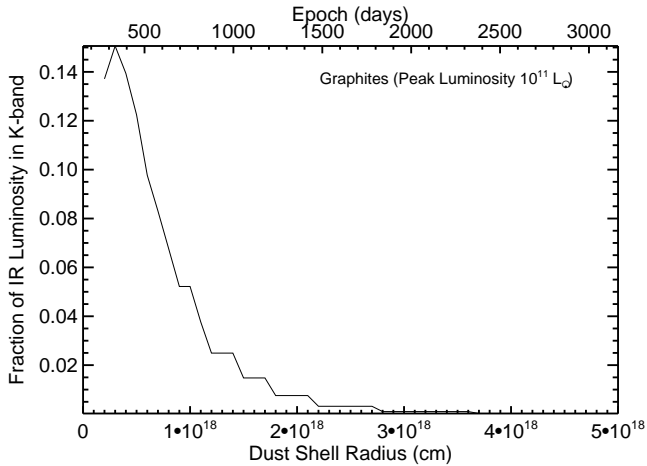}
\caption{The fraction of $K'$ to total IR luminosity as a function of dust temperature, which can be written in terms of radius (Figure \ref{fig_echotemp}) or time ($t=2R/c$), assuming a thermal light echo model and 0.1 \micron~grains.}
\label{fig_frack} 
\end{figure}

Besides the equilibrium thermal-IR emission, other potential emission sources may contribute to the $K'$-band flux at late times, including (1) the scattered-light echo at 2 \micron, (2) thermal emission from hotter, smaller grains that are not in thermal equilibrium, and (3) H$_2$ line and CO band emission.  Here we consider possible contributions from these sources.

\begin{figure}
\centering
\includegraphics[width=3.5in]{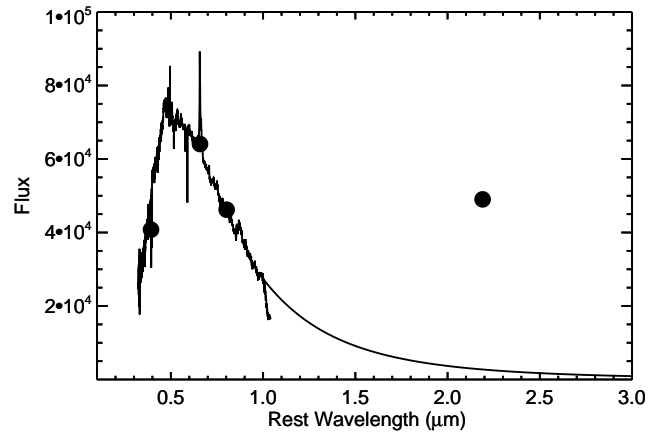}
\caption{The scattered optical light echo from Figure \ref{fig_opt_sed} extended into the IR assuming a Rayleigh-Jeans tail.  For a light-echo scenario, the $K'$-band contribution would be insignificant compared to what is observed.
\label{fig_frackscat}}
\end{figure}

{\bf Scattered-Light Echo at 2 \micron:} Figure \ref{fig_opt_sed} approximates the spectrum and SED of a scattered-light echo.  Figure \ref{fig_frackscat} goes on to extend this SED into the IR by fitting the SED with a blackbody and extending the Rayleigh-Jeans tail to the $K'$ band.  The expected fraction of the $K'$- to $R$-band flux from an unreddened scattered-light echo would be only $\sim0.02$.  From Tables \ref{tab1} and \ref{tab_nirc2_phot}, we calculate the observed ratio on day 2303 (by interpolating the $K'$-band fluxes and assuming a linear decline): $K'/R = 17.3~{\rm mag}/20.9~{\rm mag} = (9.2\times10^{-18}\, {\rm erg}~{\rm s}^{-1}~{\rm cm}^{-2}~{\rm \AA}^{-1})/(6.7\times10^{-18}\, {\rm erg}~{\rm s}^{-1}~{\rm cm}^{-2}~{\AA}^{-1}) \approx 1.38$.  We therefore rule out any significant contribution from the light echo at 2 \micron~in our models.

{\bf Small, Hot Grains:} Small grains do not radiate as blackbodies.  At a given distance from the SN, smaller grains will therefore be hotter than the equilibrium blackbody temperature exhibited by larger grains.  \citet{temim13} show in their Figure 4, however, that the dust grain temperature peaks and plateaus for dust grain sizes $a < 0.1$~\micron~(assuming a constant heating source).  Our models in Section \ref{sec:peak} already assume grain sizes $a = 0.1$~\micron~and include the associated absorption and emission coefficients (see details in \citealt{fox10}).  The modeled fraction of the IR luminosity emitted in $K'$-band flux (Figure \ref{fig_frack}) therefore represents only an upper limit.

{\bf H$_2$ line and CO band emission:} Finally, we note that additional $K'$ emission may originate from either H$_2$~line or CO band emission, but we assume the contribution is negligible in a broadband filter and do not consider these contributions in our models.

\subsection{Energy Budget}
\label{sec_budget}

The integrated optical energy output from the SN photosphere throughout the first $\sim200$ days is $\sim2.5\times10^{51}$~erg \citep{smith10}.  Assuming that the IR luminosity peaks in the $K'$ band and a constant $K'$ luminosity of $2 \times 10^8$~\lsolar~for 600 days, \citet{miller10} estimate the total emitted IR energy at $E_{\rm IR} \gtrsim 4 \times 10^{49}$~erg.  Making a similar assumption about the thermal emission peak wavelength and integrating over the observed $K'$ light curve in Figure \ref{fig_06gy_lc}, we calculate a similar total emitted energy through day 3000 (it turns out the assumption of a $K'$-band luminosity of $2 \times 10^8$~\lsolar~for 600 days was an overestimate).  

Figure \ref{fig_frack} shows, however, that in the light-echo model the fraction of the IR luminosity emitted in the $K'$ band is 15\% at day 400 and $<1$\%~by day 3000.  Accounting for the fractional output in $K'$, the lower limit on the total IR luminosity is actually $E_{\rm IR} \gtrsim 4 \times 10^{51}$~erg.  This calculation sets only a lower limit because the fractional $K'$-band output represents only the hottest dust shell.  The total radiated IR energy from the putative echo is therefore greater than the total SN output, which is not even possible for a case of an optically thick shell.  This energetics argument alone suggests that the IR echo argument is unfeasible at day 3000.  

\subsection{$R$-to-$K'$~Band Ratio}
\label{sec_ratio}

We also consider the colour evolution in the context of the light-echo scenario.  Specifically, we derive the ratio of the $R$ and $K'$ bands, which are both observables.  The scattered optical and thermal-IR fluxes can be approximated as a function of radius:
\begin{equation}
L_{R} (R) =  \frac{L_{\rm phot} \Delta t_{\rm phot} \tau_{\rm scat} f(\theta)}{2R/c},
\end{equation}
and
\begin{equation}
L_{K'} (R) =  \frac{L_{\rm phot} \Delta t_{\rm phot} \tau_{\rm d}}{2R/c}\times \frac{L_{K'}}{L_{\rm IR}}, 
\end{equation}

\begin{figure}
\centering
\includegraphics[width=3.5in]{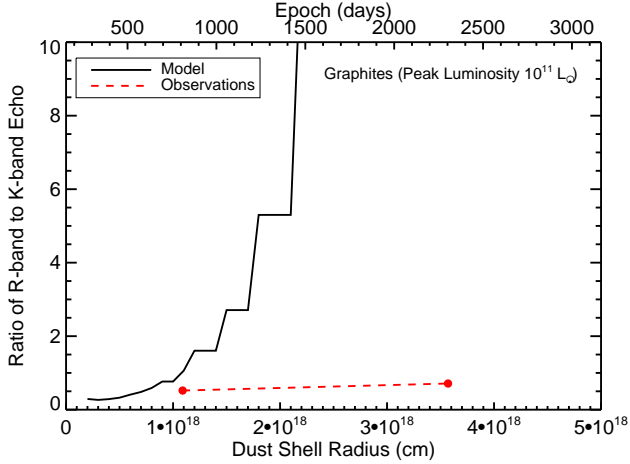}
\caption{For the light-echo model, the fraction of $K'$- to $R$-band flux as a function of dust temperature, which can be written in terms of radius (Figure \ref{fig_echotemp}) or time ($t=2R/c$)  Overplotted in red are the actual observed value from Figure \ref{fig_06gy_lc}.}
\label{fig_frac} 
\end{figure}
\noindent
where L$_{\rm phot}$~is the photosphere luminosity over a given time $\Delta t_{\rm phot}$, $\tau_{\rm scat}$~is the dust scattering coefficient, $\tau_{\rm d}$~is the dust absorption coefficient, $f(\theta)$~is the fraction of forward-scattered electrons, and $L_{K'}/L_{\rm IR}$~is the calculated IR fractional output in $K'$ (see Figure \ref{fig_frack}).  The ratio of the $R$- to $K'$-band flux versus radius can then be written as
\begin{eqnarray}
\label{eqn:ratio}
\frac{L_{R}}{L_{K'}} (R) &=& \frac{\tau_{\rm scat} f(\theta)}{\tau_{\rm d}} \times \frac{L_{IR}}{L_{\rm K'}} \\
&=& C \times \frac{L_{IR}}{L_{\rm K'}}.\nonumber
\end{eqnarray}
\noindent
While the values for $\tau_{\rm scat}$, $\tau_{\rm d}$, and $f(\theta)$~may require detailed derivations, they can all be considered constants for the purposes of this analysis, since the ratio of these values will not change as a function of the dust-shell radius.

Figure \ref{fig_frac} combines Equation \ref{eqn:ratio} with the analysis from Section \ref{sec_budget} to plot the fraction of the $R$- to $K'$-band flux as a function of radius.  As the echo shifts to shells at larger radii, the fraction of thermal echo emitted in $K'$ declines while the scattered optical light echo remains constant.  Overplotted are the measured ratio of the $R$- and $K'$-band fluxes from Figure \ref{fig_06gy_lc}.  For the first epoch, the ratio may be consistent with the light-echo scenario (depending on the values of $\tau_{\rm scat}$, $\tau_{\rm d}$, and $f(\theta)$).  The trend over $\sim1500$ days, however, does not follow the predicted shift toward a significantly larger optical flux.  These calculations again suggest that the thermal-IR echo scenario is unlikely to power the light-curve plateau, particularly by day 3000.

\subsection{Dust Temperature and Mass}
\label{sec_mass}

\citet{miller10} place a lower limit on the dust shell inner radius, $R_1 \gtrsim 1.5\times10^{18}$~cm, based on both the beginning of the near-IR echo plateau and the SN peak luminosity.  This radius should be reconsidered for two reasons.  First, a distance of $R_1 \gtrsim 1.5\times10^{18}$~cm implies a light-travel time of $>1.5$ yr.  This is inconsistent with their reported near-IR excess from a thermal echo as early as day 130.  Second, for the peak luminosity argument, \citet{miller10} assumed a dust vaporisation temperature of $T_v \approx 1000$~K, similar to that of \citet{dwek83}.  In reality, the dust vaporisation temperature for silicon and graphite dust particles is $1500 \leq T_v \leq 2000$~K \citep[e.g.,][]{gall14}.  Given the dust shell radius dependence in Equation 2 of \citet{miller10}, $R_1 \propto T_v^{-2.5}$, the calculated vaporisation radius decreases by a factor of 3--6, or $R_1 \gtrsim$ (2.5--5) $\times 10^{17}$~cm.  

This new estimate of $R_1$~changes some of the analysis and interpretation of SN 2006gy.  First, when taking into account the light-travel time, this radius is now consistent with an observed near-IR excess as early as day 130.  Second, Equation 3 of \citet{miller10} now yields a different estimate of the total dust mass in the IR-echo scenario, assuming a $r^{-2}$~wind.  If the outer dust radius is given by the light-travel time from the most recent set of Keck/NIRC2 observations (day 3024), then $R_2 \gtrsim ct/2 
= 4 \times 10^{18}$~cm = (8--16)$R_1$.  According to Equation 3 in \citet{miller10}, this increases the estimate of the total dust mass by nearly a factor of $\sim2$--3, or $M_d \approx 0.2$--0.3~\msolar.

\begin{figure}
\centering
\includegraphics[width=3.5in]{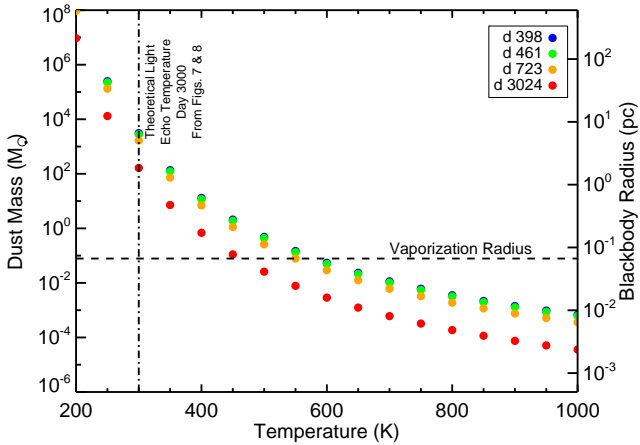}
\caption{A single $K'$-band flux can be fit as a function of the dust mass and temperature (a higher temperature requires a lower dust mass and vice versa).  The different combinations from this relationship are calculated for the measured $K'$ flux at each epoch.}
\label{fig_tempvmass} 
\end{figure}

This dust mass derived from the light-echo scenario is difficult to reconcile with the observed $K'$-band magnitudes.  Following the analysis of \citet{fox10} and \citet{fox11}, a single $K'$-band flux can be fit as a function of the dust mass and temperature (e.g., a higher temperature requires a lower dust mass and vice versa).  Figure \ref{fig_tempvmass} shows this relationship for the measured $K'$-band flux at each epoch.  

On day 3024, Figure \ref{fig_echotemp} indicates that for the light-echo model, the inner dust-shell radius, $R \approx 5 \times 10^{18}$~cm, will have have a temperature of $<$300 K.  The hottest dust serves as a useful lower limit on the required dust mass because colder dust would contribute even less flux to the measured $K'$-band flux.  Figure \ref{fig_tempvmass} reveals that to recreate the observed $K'$ flux with 300 K dust would require $>10^2$~\msolar.  This mass is significantly greater than the dust mass derived above from the light-echo equations.  Furthermore, such a large dust mass would require $>10^4$~\msolar~of gas (assuming typical gas-to-dust mass ratios of $\sim100$).  Taken all together, these results offer further evidence that the thermal-echo scenario is not likely the dominant mechanism powering the late-time IR plateau of SN 2006gy by day 3000.

\section{IR Emission From CSM Interaction?}
\label{sec:csm}

Other physical scenarios may explain the late-time IR plateau of SN 2006gy, including radioactive decay from a pair-instability SN, new dust formation, collisional heating, and radiative heating from CSM interaction.  \citet{smith08gy} and \citet{miller10} rule out most of these scenarios with earlier data.  Specifically, \citet{smith08gy} argue against CSM interaction at $\lesssim$800 days given the absence of broad or intermediate H$\alpha$ emission in a day 364 2D spectrum (but see Section \ref{sec_stis_spectrum} below), weak X-rays, and no radio detection.  They also rule out the possibility of dust obscuration, since the calculated dust mass would be insufficient to hide the expected H$\alpha$~luminosity.  These calculations assume a dust temperature of $T_d = 1300$~K so that the energy distribution peaks in the $K'$ band.  

While the light echo may have dominated at these earlier epochs, such assumptions are no longer valid by day 3000.  Here we consider the possibility that CSM interaction dominates the IR light curve at these later epochs with new assumptions and data.  In this scenario, a pre-existing dust shell is radiatively heated by X-ray and/or optical emission generated by ongoing CSM interaction \citep{fox11,fox13}.  Throughout the analysis, we assume a spherically symmetric distribution of dust.

\subsection{Dust Temperature and Mass}
\label{sec_mass_csm}

Unlike the thermal-echo scenario, where only a fraction of each dust shell emits at a given epoch, the observed IR flux in the CSM interaction scenario corresponds to the entire dust shell (as long as the light crossing time is shorter than the duration of CSM interaction).  Assuming spherical symmetry, the IR flux can therefore be converted into a minimum dust shell blackbody radius,
$r_{\rm bb} = (L_{\rm IR}/4 \pi \sigma T_d^4)^{1/2}$.  
Figure \ref{fig_tempvmass} plots the blackbody radius corresponding to each combination of dust mass and temperature used to fit the $K'$ fluxes at each epoch (see Section \ref{sec_mass}), although a number of degeneracies remain.

Independent constraints can break these degeneracies in Figure \ref{fig_tempvmass}.  Figure \ref{fig_echotemp} demonstrates that for a peak luminosity of $\sim10^{11}$~\lsolar, the dust vaporisation radius (assuming a vaporisation temperature of $T_d \approx 1500$~K) is $\sim2 \times 10^{17}$~cm~$= 0.067$~pc.  For $r_{\rm bb} = 0.067$~pc, Figure \ref{fig_tempvmass} shows a corresponding dust-shell temperature and mass on day 3024 of $\sim450$~K and $8\times10^{-2}$~\msolar, respectively.  Figures \ref{fig_echotemp} and \ref{fig_frack} highlight that a temperature $T_d \approx 450$~K corresponds to a fractional $K'$-band output of $\sim2$\%.  (These calculations assume a dust-shell radius equal to the vaporisation radius, which requires a constant pre-SN mass loss.  Of course, the pre-SN mass loss may not have been constant and the dust-shell radius may be larger.  We consider this possibility later.)

\subsection{Energy Budget}
\label{sec_budget_csm}

Accounting for this fractional IR flux in $K'$, the total emitted IR and optical energy throughout the plateau phase is $E_{\rm IR} \gtrsim 2 \times 10^{50}$~erg and $E_{\rm opt} \sim2\times10^{49}$~erg (see Section \ref{sec_budget}).  For the CSM interaction scenario, the thermal-IR emission results when the dust shell absorbs and reradiates the optical flux generated by the shocks.  The implied optical depth is therefore
\begin{eqnarray}
\tau_d & \approx & -1 \times {\rm ln}\bigg(\frac{E_{\rm opt}}{E_{\rm opt}+E_{\rm IR}}\bigg) \\
& = & 2.3\,. \nonumber
\end{eqnarray}

\begin{figure*}
\centering
\hspace{-0.5in}
\includegraphics[width=7.7in]{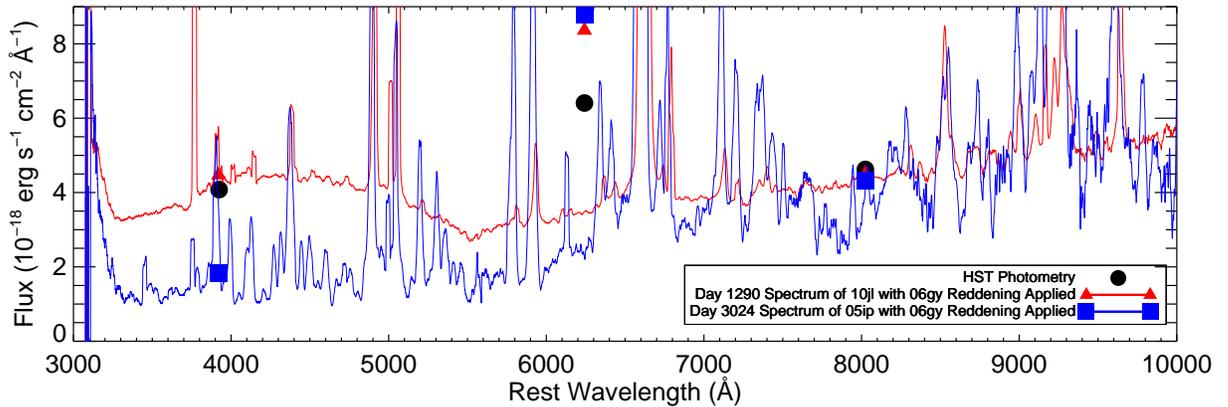}
\caption{The optical SED of SN 2006gy on day 2379, which corresponds to the only epoch of F390W observations (similar to Figure \ref{fig_opt_sed}).  Overplotted is the day 1290 spectrum of SN 2010jl (red) and the day 3024 spectrum of SN 2005ip (red), both which show evidence for CSM interaction.  Both spectra are scaled to the F814W photometry.  Strong H$\alpha$~for both SNe 2005ip and 2010jl results in a poor match with the observed F625W emission from SN 2006gy.}
\label{fig_opt_sed2} 
\end{figure*}

Assuming most of the optical energy is emitted in the $R$ band, the extinction can be estimated as $A_R \approx \tau_d = 2.3$, which is slightly larger than estimated by \citet{smith08gy}.  A slightly warmer dust shell or additional visible radiation outside of $R$ would decrease the value of $A_R$~toward the value of 1.5 measured by \citet{smith08gy}.  

The integrated optical energy output from the SN photosphere throughout the first $\sim200$ days is $\sim2.5\times10^{51}$~erg (see Section \ref{sec_budget}).  Assuming CSM interaction dominates at later epochs, the total radiated energy emitted from the SN throughout day 3000 is therefore $E_{\rm tot} \approx 2.5\times10^{51} + 4 \times 10^{50} + 2\times10^{49} = 2.9\times10^{51}$~erg.  This number is larger than the radiated energy output by most SNe, but consistent with many SLSN, especially considering observations of SN 2006gy extend all the way through day 3000 \citep[see Table 1 in][]{gal-yam12}.

\subsection{$R$-to-$K'$~Band Ratio}
\label{sec_ratio_csm}

Figure \ref{fig_frac} plots the ratio of the observed $R$- and $K'$-band fluxes.  The ratio remains nearly constant throughout the extent of the observations.  While this trend is not consistent with the expected colour evolution of a light echo, it is expected for the CSM interaction scenario.  If the shock remains interior to a relatively static dust shell and the optical depth stays constant, the ratio of the absorbed to emitted flux will also be constant.  The $K'$ flux may be expected to drop slightly as dust is destroyed by the forward shock, leaving only dust at larger radii.

\subsection{Can CSM Interaction Also Explain the Optical Emission?}

Figure \ref{fig_opt_sed} already shows that the observed SED of SN 2006gy on day 2379 is consistent with a scattered-light echo.  We now consider the possibility that the optical SED may also be consistent with CSM interaction.  Figure \ref{fig_opt_sed2} again plots the observed SED of SN 2006gy on day 2379.  This time, however, the figure goes on to compare the photometry to the late-time Keck spectra obtained of both SNe 2005ip (day 3024) and 2010jl (day 1290), two Type IIn explosions that are known to exhibit significant CSM interaction at very late epochs.  To make for a fair comparison, both spectra are reddened by a similar amount as SN 2006gy ($E(B - V) = 0.72$ mag).  Since we expect minimal contributions of CSM interaction at redder wavelengths, the spectra are scaled so that their synthetic photometry matches the F814W photometry of SN 2006gy.

The synthetic photometry of the two spectra is comparable with the F390W and F814W photometry of SN 2006gy (SN 2010jl more so than SN 2005ip), but it significantly overestimates the F625W photometry.  The most likely explanation for this is that both SNe 2010jl and 2005ip have strong H$\alpha$ emission lines that are consistent with strong and ongoing CSM interaction \citep{smith09ip,smith12jl}, while SN 2006gy does not exhibit prominent H$\alpha$.  We search more carefully for H$\alpha$~in the day 3001 STIS spectrum of SN 2006gy in Section \ref{sec_stis_spectrum} below.

Since the SED and spectral shapes are at least somewhat consistent, however, this comparison raises the question of whether some other late-time spectrum with less CSM interaction may offer a better fit (see, for example, \citealt{fox13}).  While this paper does not compare an exhaustive set of late-time spectra exhibiting CSM interaction, we point out that it would be difficult (if not unphysical) to produce significant line emission in the F390W bandpass (e.g., ``blue pseudo-continuum'') without also generating strong H$\alpha$~emission in the F625W filter.  Furthermore, these late-time spectra may be misleading because the SNe can be in a star cluster, have an optical echo of their own, or have bad subtraction of galaxy light, which is common when the SNe are faint.  We therefore conclude that the optical emission from SN 2006gy is most likely dominated by a scattered-light echo, although CSM interaction may contribute a small fraction of the observed emission.

\subsection{STIS Spectrum and the H$\alpha$~Flux}
\label{sec_stis_spectrum}

\begin{figure}
\centering
\includegraphics[width=3.5in]{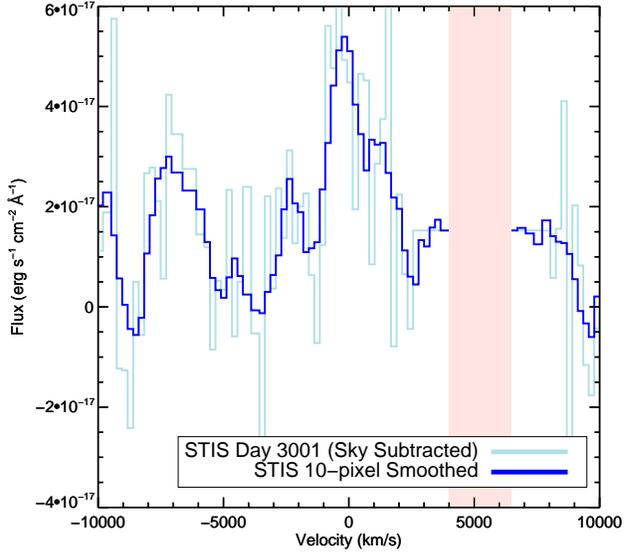}
\caption{Velocity profile of the H$\alpha$~line.  Although noisy, the H$\alpha$~line is broad (FWHM = $2000\pm200$ \kms).  At the very least, this rules out the possibility of narrow emission from an underlying H~II region dominating the line flux \citep[e.g.,][]{fox13}.  A light-red bar signifies pixels flagged for having bad data quality.}
\label{fig_stisvel} 
\end{figure}

Figure \ref{fig_stis_spectrum} plots the STIS spectrum corrected for extinction assuming a reddening $E(B - V) = 0.72$ mag.  While the day 3001 STIS spectrum is noisy, we can identify the H$\alpha$~line, particularly in the 2D image (see Figure \ref{fig_2d}).  Figure \ref{fig_stisvel} further confirms the presence of a broad, full width at half-maximum intensity (FWHM) $2000\pm200$ \kms\ profile.  This velocity is somewhat consistent with the speed of the post-shock gas ($\sim4000$ \kms; \citealt{smith08gy}), which shouldn't decelerate significantly if the shock is plowing into a $r^{-2}$~wind.  No other lines can be definitively identified.

We estimate the total output from the H$\alpha$~line by integrating over the H$\alpha$~line profile after subtracting a continuum of $\sim5\times10^{-18}$~erg s$^{-1}$~cm$^{-2}$~\AA$^{-1}$.  The integrated flux of the line before correcting for extinction is $(6.5 \pm 0.7) \times10^{39} $~\ergs\, which is about an order of magnitude larger than the upper limit placed on it by \citet{smith08gy}.  Some of this emission, however, originates in the scattered-light echo (see Section \ref{sec:optical}).  We approximate the contribution from CSM interaction by subtracting off the excess SN flux in the F625W filter compared to the synthetic photometry of the scattered-light-echo model in Figure \ref{fig_opt_sed}, which comes out to $\sim4\times10^{38}$~\ergs, or 5\% of the F625W flux (which is also within the error bar).  We point out that the STIS spectrum was obtained nearly two years after the SED constructed in Figure \ref{fig_opt_sed}, so the relative contributions of the scattered echo and CSM interaction may have changed.

Despite the identification of the broad H$\alpha$~line, we consider the possibility that there is also some narrow-line contribution from an underlying H~II~region \citep[e.g.,][]{fox13}.  For a comparison, we consider the H~II region of the Carina Nebula, the host to $\eta$ Carinae.  While large when considering our local Milky Way neighbourhood, the Nebula is small relative to other known massive star forming complexes, consistent with the home environment of the likely massive progenitor to SN 2006gy.  At the distance of SN 2006gy, the whole Nebula would span only $\sim0.2$\arcsec, which would be spatially unresolved in these spectra.  The Carina Nebula has an integrated H$\alpha$~luminosity of $10^5$~\lsolar~\citep{smithbrooks07}, which would leave nearly none of the remaining available H$\alpha$~line flux to originate from CSM interaction.  This flux is within the limits independently calculated by \citet{smith10} and \citet{agnoletto09}. 

\subsection{Powering the IR Emission}
\label{sec_powering}

Although we do not have a constraint on the dust mass or shell radius, Figure \ref{fig_tempvmass} shows that the observed $K'$-band flux on day 3024 can be generated with $\sim450$ K dust in a spherically symmetric shell at the vaporisation radius.  Figure 8(b) from \citet{fox10} calculates that 0.1 \micron~graphite dust at this temperature and radius requires a powering source with a luminosity of $\sim10^9$~\lsolar$ = 3.9\times10^{42}$~\ergs.  Taking a CSM luminosity contribution of even $5\times10^{38}$~\ergs (see Section \ref{sec_stis_spectrum}) and an optical depth of $\tau_d \approx A_R \approx 2.3$~yields a total H$\alpha$~luminosity arising from CSM interaction of $\sim5 \times 10^{39}$~\ergs.  While this emission alone is not sufficient to heat the dust to the observed temperature, the bulk of emission from CSM interaction most likely arises in the X-rays and UV, where there are many emission lines at wavelengths where dust grains absorb very efficiently.  Such an effect is commonly observed in supernova remnants \citep[e.g.,][]{temim12} and other interacting SNe \citep[e.g.,][]{fransson02}.  In fact, \citet{chevalier94} show that the conversion efficiency from X-rays to H$\alpha$~is only $\sim1$\%, suggesting X-ray luminosities of $\sim5 \times 10^{41}$~\ergs.

While we do not have any X-ray observations available at the time of this study, we do have nearly contemporaneous UV observations (see Table \ref{tab1}).  Taking into account an optical depth of $\tau_d \approx A_R \approx 2.3$~yields upper limits on the near-UV and far-UV fluxes of $5.5\times10^{40}$~and $6.8\times10^{40}$~\ergs~, respectively, for a total UV output of $\sim2.3\times10^{41}$~\ergs.  

While the UV output is still not sufficient to power the $K'$-band emission at the vaporisation radius, we note several caveats.  Again, we stress that a significant portion of the flux arising from CSM interaction is likely emitted in the X-rays, for which we do not have data.  Furthermore, the vaporisation radius sets only a lower limit on the dust-shell radius.  Figure \ref{fig_tempvmass} shows that a shell with a radius just a factor of 3 larger can yield the same $K'$-band flux with a more massive dust shell that is $>100$~K colder, which in turn requires $\sim10^8$~\lsolar$=3.9\times10^{41}$~\ergs\ to heat it.  This scenario is consistent with the upper limits provided by the UV observations and X-ray estimates (assuming a 1\%~conversion efficiency).

\section{Discussion and Conclusion}
\label{sec:discussion}

In this article, we present new observations of the 3000 day plateau in SN 2006gy, including Keck/NIRC2-AO-LGS $K'$-band photometry on day 3024 and an {\it HST}/STIS spectrum on day 3001.  After examining the energetics, colour evolution, dust mass, and presence of broad H$\alpha$~in the optical spectrum, we find that the new data are consistent with the presence of both a scattered optical light echo and IR emission from dust radiatively heated by X-rays and UV emission from ongoing CSM interaction.

Even considering the 3000 day plateau, the total radiated energy output calculated in Section \ref{sec_budget_csm} is only $E_{\rm tot} = 2.9\times10^{51}$~erg, which is not significantly greater than the total energy emitted by the SN during the first $\sim$220 days and still within the $5 \times 10^{51}$~erg of kinetic energy calculated by \citet{smith10}.  The derived energy budget therefore does not contradict the CSM interaction model for SN 2006gy that invokes a large CSM formed by a very massive progenitor star \citep{smith10}.

Assuming that the CSM interaction scenario provides the dominant power source for the day 3000 IR dust emission, the total dust-shell mass can be tied to the progenitor's total mass-loss rate, 
\begin{eqnarray}
\label{eqn:outer}
\mdot_{\rm outer} & = & \frac{M_{\rm d}}{Z_{\rm d} \Delta r} v_{\rm w} \nonumber \\ 
& = & \frac{3}{4} \Big(\frac{M_{\rm d}}{\rm M_{\odot}}\Big) \Big(\frac{v_{\rm w}}{120~\rm km~s^{-1}}\Big) \nonumber\\
&\times& \Big(\frac{5 \times 10^{16}~\rm cm}{r}\Big) \Big(\frac{r}{\Delta r}\Big)~ {\rm M}_{\odot}~{\rm yr}^{-1},
\end{eqnarray}
for a dust-to-gas mass ratio $Z_{\rm d} = M_{\rm d}/M_{\rm g} \approx 0.01$, and a progenitor wind speed $v_{\rm w}$.  The relatively narrow lines observed in SNe~IIn originate in the slow pre-shocked CSM and can be used to approximate the progenitor wind speed.  The narrow lines observed in SN 2006gy have widths corresponding to $v_{\rm w} =$ 200 \kms\ \citep{smith08gy}.  Assuming a thin shell ${\Delta r}/r = 1/10$, dust-shell radius $r = 3\times10^{17}$~cm (see Section \ref{sec_mass_csm}), and dust mass $M_d = 0.1$~\msolar, the approximate mass-loss rate to produce the observed dust shell is $\mdot_{\rm outer} \approx 0.2$~\ml.  A larger shell radius would require a larger dust mass (see Figure \ref{fig_tempvmass}), but these nearly cancel each other out and therefore do not change the calculated mass-loss rate significantly.

The optical and/or UV/X-ray emission generated by CSM interaction can served as a tracer for the mass loss at the inner, shock radius.  Assuming a density $\propto r^{-2}$~wind profile, the rate can be written as a function of the optical/X-ray luminosity, progenitor wind speed, and shock velocity \citep[e.g.,][]{chugai94, smith09ip}:
\begin{eqnarray}
\label{eqn:inner}
\mdot_{\rm inner} & = & \frac{2 v_{\rm w}}{\epsilon v_{\rm s}^3}L_{\rm opt/UV/X}, \nonumber \\
 & = & 2.1 \times 10^{-4} \Big(\frac{L_{\rm opt/UV/X}}{3~\times 10^{41}~ \ergs}\Big) \nonumber\\
  &\times&\Big(\frac{\epsilon}{0.5}\Big)^{-1} \Big(\frac{v_{\rm w}}{120~\kms}\Big) \nonumber\\
  &\times& \Big( \frac{v_{\rm s}}{10^4~\kms}\Big)^{-3} {\rm M_{\odot}~yr^{-1}},
\end{eqnarray}
where $\epsilon < 1$~is the efficiency of converting shock kinetic energy into light.  We assume a value $\epsilon \approx 0.1$, although the conversion efficiency can vary with wind density and shock speed.  Depending on the extinction correction we apply, a late-time optical luminosity of $L_R \approx 10^{41}$~\ergs, wind speed $v_{\rm w} = 200$ \kms, shock velocity $v_{\rm s} =$ 4000 \kms~\citep{smith08gy}, and conversion efficiency $\epsilon=0.1$ correspond to a mass-loss rate $\mdot_{\rm inner} \approx 10^{-2}$~\ml.  

Equations \ref{eqn:outer} and \ref{eqn:inner} yield only order-of-magnitude approximations, but it appears that the inner CSM mass-loss rate is an order of magnitude smaller than the mass-loss rate for the outer dust shell.  The dust shell was likely formed during a period of increased, nonsteady mass loss.  Furthermore, assuming spherical symmetry, the total implied dust mass from Figure \ref{fig_tempvmass} is $\gtrsim 8\times 10^{-2}$~\msolar.  While this number is comparable to that of many other SNe~IIn \citep{fox11,fox13}, it represents some of the most massive shells observed around SNe~IIn.  

As a comparison, consider that the total warm dust mass of SN 2006gy is nearly as large as the total cold ($\sim20$~K) dust mass observed in the extremely dusty SN 1987A, which was detected only due to its nearby distance from Earth \citep{matsuura11, Indebetouw14, matsuura15}.  The implication is that much larger dust reservoirs may be hiding in SN 2006gy.  If this much dust forms in other SLSNe and can survive the forward and/or reverse shocks, it may be able to account for the cosmic dust budget \citep[see][and references therein]{gall14}.  The large dust shell and extremely dense CSM may also explain the suppressed X-ray and radio emission, an effect observed to lesser degrees in other SNe~IIn \citep[e.g.,][]{vandyk96,fox09}.

The total dust mass of $\gtrsim10^{-1}$~\msolar~derived from CSM interaction indicates a total shell mass of $\sim$10~\msolar, assuming standard gas-to-dust mass ratios.  As noted by \citet{smith08gy} and \citet{miller10}, such a large envelope suggests a very massive progenitor star.  These numbers are surprisingly close to those derived by \citet{smith08gy} and \citet{miller10}, which is probably not a mere coincidence.  More likely, the same dust shell observed as a thermal echo by \citet{smith08gy} and \citet{miller10} is increasingly heated by CSM interaction.  

The derived radius of the dust shell in this paper, however, differs from that of \citet{smith08gy} and \citet{miller10}.  For the CSM interaction scenario, the shell resides nearly a factor of 10 closer, which would suggest the observed dust shell was formed in an outburst that occurred as recently as just a few hundred years prior to the explosion.  The time frame of the outbursts may offer important constraints on the progenitor mass-loss mechanisms.  As the forward shock overtakes this shell of material, we expect the SN luminosity to increase for a period of time consistent with the width of the shell of material.  Continued monitoring of SN 2006gy will help to constrain the dust-shell properties.

Overall, this paper highlights the significance of multi-wavelength observations, even if limited in wavelength range and sensitivity.  Future observations of interacting SNe should cover more wavelengths (e.g., X-ray, mid-IR, and radio) over more epochs.  Ongoing observations of SN 2006gy will ultimately constrain the precise extent of this material, the presence of additional shells, and the possibility of alternative, more exotic energy sources, such as a magnetar.

\vspace{0.5in}

This work is based on observations made with the NASA/ESA {\it Hubble Space Telescope}, obtained from the Space Telescope Science Institute (STScI), which is operated by the Association of Universities for Research in Astronomy (AURA), Inc., under NASA contract NAS5-26555. We are grateful to the STScI Help Desk for their assistance with the {\it HST}~data. Some of the data presented herein were obtained at the W.~M. Keck Observatory, which is operated as a scientific partnership among the California Institute of Technology, the University of California, and NASA; the observatory was made possible by the generous financial support of the W.~M. Keck Foundation.  The Keck observations were made possible by the ToO program.  We thank the staff of the Keck Observatory for their assistance with the observations, as well as efforts by Sam Ragland and Mark Morris.  Melissa L. Graham and WeiKang Zheng helped obtain and reduce the Keck spectra.  The authors wish to recognise and acknowledge the very significant cultural role and reverence that the summit of Mauna Kea has always had within the indigenous Hawaiian community.  We are most fortunate to have the opportunity to conduct observations from this mountain.  

Financial support for O.D.F. was provided by NASA through grant GO-13287 from STScI. A.V.F. and his group acknowledge generous financial assistance from the Christopher R. Redlich Fund, the TABASGO Foundation, and NSF grant AST-1211916.  The research by S.M.A is supported by the U.S. Department of Energy through the Lawrence Livermore National Laboratory under Contract DE-AC52-07NA27344.

\bibliographystyle{apj2}
\bibliography{references}

\begin{thebibliography}{}
\expandafter\ifx\csname natexlab\endcsname\relax\def\natexlab#1{#1}\fi

\bibitem[{Agnoletto {et~al.}(2009)Agnoletto, Benetti, Cappellaro, Zampieri,
  Turatto, Mazzali, Pastorello, Valle, Bufano, Harutyunyan, Navasardyan,
  Elias-Rosa, Taubenberger, Spiro, \& Valenti}]{agnoletto09}
Agnoletto, I., Benetti, S., Cappellaro, E., {et~al.} 2009, The Astrophysical
  Journal, 691, 1348

\bibitem[{Bode \& Evans(1980)}]{bode80}
Bode, M.~F., \& Evans, A. 1980, A{\&}A, 89, 158

\bibitem[{Cardelli {et~al.}(1989)Cardelli, Clayton, \& Mathis}]{cardelli89}
Cardelli, J.~A., Clayton, G.~C., \& Mathis, J.~S. 1989, ApJ, 345, 245

\bibitem[{Chevalier(1986)}]{chevalier86}
Chevalier, R.~A. 1986, ApJ, 308, 225

\bibitem[{Chevalier \& Fransson(1994)}]{chevalier94}
Chevalier, R.~A., \& Fransson, C. 1994, Astrophysical Journal, 420, 268

\bibitem[{Chugai \& Danziger(1994)}]{chugai94}
Chugai, N.~N., \& Danziger, I.~J. 1994, MNRAS, 268, 173

\bibitem[{Dolphin(2000)}]{dolphin00}
Dolphin, A.~E. 2000, PASP, 112, 1383

\bibitem[{Dwek(1983)}]{dwek83}
Dwek, E. 1983, ApJ, 274, 175

\bibitem[{Filippenko(1982)}]{filippenko82}
Filippenko, A.~V. 1982, PASP, 94, 715

\bibitem[{Filippenko(1997)}]{filippenko97}
---. 1997, ARA\&A, 35, 309

\bibitem[{Foley {et~al.}(2003)Foley, Papenkova, Swift, Filippenko, Li, Mazzali,
  Chornock, Leonard, \& van Dyk}]{foley03}
Foley, R.~J., Papenkova, M.~S., Swift, B.~J., {et~al.} 2003, PASP, 115, 1220

\bibitem[{Fox {et~al.}(2010)Fox, Chevalier, Dwek, Skrutskie, Sugerman, \&
  Leisenring}]{fox10}
Fox, O.~D., Chevalier, R.~A., Dwek, E., {et~al.} 2010, ApJ, 725, 1768

\bibitem[{Fox {et~al.}(2013)Fox, Filippenko, Skrutskie, Silverman,
  Ganeshalingam, Cenko, \& Clubb}]{fox13}
Fox, O.~D., Filippenko, A.~V., Skrutskie, M.~F., {et~al.} 2013, AJ, 146, 2

\bibitem[{Fox {et~al.}(2009)Fox, Skrutskie, Chevalier, Kanneganti, Park,
  Wilson, Nelson, Amirhadji, Crump, Hoeft, Provence, Sargeant, Sop, Tea,
  Thomas, \& Woolard}]{fox09}
Fox, O.~D., Skrutskie, M.~F., Chevalier, R.~A., {et~al.} 2009, ApJ, 691, 650

\bibitem[{Fox {et~al.}(2011)Fox, Chevalier, Skrutskie, Soderberg, Filippenko,
  Ganeshalingam, Silverman, Smith, \& Steele}]{fox11}
Fox, O.~D., Chevalier, R.~A., Skrutskie, M.~F., {et~al.} 2011, ApJ, 741, 7

\bibitem[{Fransson {et~al.}(2002)Fransson, Chevalier, Filippenko, Leibundgut,
  Barth, Fesen, Kirshner, Leonard, Li, Lundqvist, Sollerman, \& van
  Dyk}]{fransson02}
Fransson, C., Chevalier, R.~A., Filippenko, A.~V., {et~al.} 2002, The
  Astrophysical Journal, 572, 350

\bibitem[{Gal-Yam(2012)}]{gal-yam12}
Gal-Yam, A. 2012, Science, 337, 927

\bibitem[{Gall {et~al.}(2014)Gall, Hjorth, Watson, Dwek, Maund, Fox, Leloudas,
  Malesani, \& Day-Jones}]{gall14}
Gall, C., Hjorth, J., Watson, D., {et~al.} 2014, Nature, 511, 326

\bibitem[{Horne(1986)}]{horne86}
Horne, K. 1986, PASP, 98, 609

\bibitem[{Indebetouw {et~al.}(2014)Indebetouw, Matsuura, Dwek, Zanardo, Barlow,
  Baes, Bouchet, Burrows, Chevalier, Clayton, Fransson, Gaensler, Kirshner,
  Laki{\'c}evi{\'c}, Long, Lundqvist, Mart{\'\i}-Vidal, Marcaide, McCray,
  Meixner, Ng, Park, Sonneborn, Staveley-Smith, Vlahakis, \& van
  Loon}]{Indebetouw14}
Indebetouw, R., Matsuura, M., Dwek, E., {et~al.} 2014, ApJL, 782, L2

\bibitem[{Matheson {et~al.}(2000)Matheson, Filippenko, Ho, Barth, \&
  Leonard}]{matheson00}
Matheson, T., Filippenko, A.~V., Ho, L.~C., Barth, A.~J., \& Leonard, D.~C.
  2000, AJ, 120, 1499

\bibitem[{Matsuura {et~al.}(2011)Matsuura, Dwek, Meixner, Otsuka, Babler,
  Barlow, Roman-Duval, Engelbracht, Sandstrom, Laki{\'c}evi{\'c}, Loon,
  Sonneborn, Clayton, Long, Lundqvist, Nozawa, Gordon, Hony, Panuzzo, Okumura,
  Misselt, Montiel, \& Sauvage}]{matsuura11}
Matsuura, M., Dwek, E., Meixner, M., {et~al.} 2011, Science, 333, 1258

\bibitem[{Matsuura {et~al.}(2015)Matsuura, Dwek, Barlow, Babler, Baes, Meixner,
  Cernicharo, Clayton, Dunne, Fransson, Fritz, Gear, Gomez, Groenewegen,
  Indebetouw, Ivison, Jerkstrand, Lebouteiller, Lim, Lundqvist, Pearson,
  Roman-Duval, Royer, Staveley-Smith, Swinyard, van Hoof, Loon, Verstappen,
  Wesson, Zanardo, Blommaert, Decin, Reach, Sonneborn, de~Steene, \&
  Yates}]{matsuura15}
Matsuura, M., Dwek, E., Barlow, M.~J., {et~al.} 2015, ApJ, 800, 50

\bibitem[{Miller {et~al.}(2010)Miller, Smith, Li, Bloom, Chornock, Filippenko,
  \& Prochaska}]{miller10}
Miller, A.~A., Smith, N., Li, W., {et~al.} 2010, AJ, 139, 2218

\bibitem[{Nomoto {et~al.}(2007)Nomoto, Tominaga, Tanaka, Maeda, \&
  Umeda}]{nomoto07}
Nomoto, K., Tominaga, N., Tanaka, M., Maeda, K., \& Umeda, H. 2007, SUPERNOVA
  1987A: 20 YEARS AFTER: Supernovae and Gamma-Ray Bursters. AIP Conference
  Proceedings, 937, 412

\bibitem[{Ofek {et~al.}(2007)Ofek, Cameron, Kasliwal, Gal-Yam, Rau, Kulkarni,
  Frail, Chandra, Cenko, Soderberg, \& Immler}]{ofek07}
Ofek, E.~O., Cameron, P.~B., Kasliwal, M.~M., {et~al.} 2007, ApJ, 659, L13

\bibitem[{Oke {et~al.}(1995)Oke, Cohen, Carr, Cromer, Dingizian, Harris,
  Labrecque, Lucinio, Schaal, Epps, \& Miller}]{oke95}
Oke, J.~B., Cohen, J.~G., Carr, M., {et~al.} 1995, PASP, 107, 375

\bibitem[{Peng {et~al.}(2002)Peng, Ho, Impey, \& Rix}]{peng02}
Peng, C.~Y., Ho, L.~C., Impey, C.~D., \& Rix, H.-W. 2002, AJ, 124, 266

\bibitem[{Peng {et~al.}(2010)Peng, Ho, Impey, \& Rix}]{peng10}
---. 2010, AJ, 139, 2097

\bibitem[{Prieto {et~al.}(2006)Prieto, Garnavich, Chronister, \&
  Connick}]{prieto06}
Prieto, J.~L., Garnavich, P., Chronister, A., \& Connick, P. 2006, CBET, 648, 1

\bibitem[{Quimby(2006)}]{quimby06}
Quimby, R. 2006, CBET, 644, 1

\bibitem[{S{\'e}rsic(1963)}]{sersic63}
S{\'e}rsic, J.~L. 1963, Boletin de la Asociacion Argentina de Astronomia, 6, 41

\bibitem[{Silverman {et~al.}(2012)Silverman, Foley, Filippenko, Ganeshalingam,
  Barth, Chornock, Griffith, Kong, Lee, Leonard, Matheson, Miller, Steele,
  Barris, Bloom, Cobb, Coil, Desroches, Gates, Ho, Jha, Kandrashoff, Li,
  Mandel, Modjaz, Moore, Mostardi, Papenkova, Park, Perley, Poznanski, Reuter,
  Scala, Serduke, Shields, Swift, Tonry, van Dyk, Wang, \&
  Wong}]{silverman12bsnip1}
Silverman, J.~M., Foley, R.~J., Filippenko, A.~V., {et~al.} 2012, MNRAS, 425,
  1789

\bibitem[{Skrutskie {et~al.}(2006)Skrutskie, Cutri, Stiening, Weinberg,
  Schneider, Carpenter, Beichman, Capps, Chester, Elias, Huchra, Liebert,
  Lonsdale, Monet, Price, Seitzer, Jarrett, Kirkpatrick, Gizis, Howard, Evans,
  Fowler, Fullmer, Hurt, Light, Kopan, Marsh, McCallon, Tam, Dyk, \&
  Wheelock}]{skrutskie06}
Skrutskie, M.~F., Cutri, R.~M., Stiening, R., {et~al.} 2006, AJ, 131, 1163

\bibitem[{Smith \& Brooks(2007)}]{smithbrooks07}
Smith, N., \& Brooks, K.~J. 2007, Monthly Notices of the Royal Astronomical
  Society, 379, 1279

\bibitem[{Smith {et~al.}(2010)Smith, Chornock, Silverman, Filippenko, \&
  Foley}]{smith10}
Smith, N., Chornock, R., Silverman, J.~M., Filippenko, A.~V., \& Foley, R.~J.
  2010, ApJ, 709, 856

\bibitem[{Smith {et~al.}(2012)Smith, Silverman, Filippenko, Cooper, Matheson,
  Bian, Weiner, \& Comerford}]{smith12jl}
Smith, N., Silverman, J.~M., Filippenko, A.~V., {et~al.} 2012, ApJ, 143, 17

\bibitem[{Smith {et~al.}(2007)Smith, Li, Foley, Wheeler, Pooley, Chornock,
  Filippenko, Silverman, Quimby, Bloom, \& Hansen}]{smith07}
Smith, N., Li, W., Foley, R.~J., {et~al.} 2007, ApJ, 666, 1116

\bibitem[{Smith {et~al.}(2008)Smith, Foley, Bloom, Li, Filippenko, Gavazzi,
  Ghez, Konopacky, Malkan, Marshall, Pooley, Treu, \& Woo}]{smith08gy}
Smith, N., Foley, R.~J., Bloom, J.~S., {et~al.} 2008, ApJ, 686, 485

\bibitem[{Smith {et~al.}(2009)Smith, Silverman, Chornock, Filippenko, Wang, Li,
  Ganeshalingam, Foley, Rex, \& Steele}]{smith09ip}
Smith, N., Silverman, J.~M., Chornock, R., {et~al.} 2009, ApJ, 695, 1334

\bibitem[{Temim \& Dwek(2013)}]{temim13}
Temim, T., \& Dwek, E. 2013, ApJ, 774, 8

\bibitem[{Temim {et~al.}(2012)Temim, Slane, Arendt, \& Dwek}]{temim12}
Temim, T., Slane, P., Arendt, R.~G., \& Dwek, E. 2012, The Astrophysical
  Journal, 745, 46

\bibitem[{{Van Dyk} {et~al.}(1996){Van Dyk}, Weiler, Sramek, Schlegel,
  Filippenko, Panagia, \& Leibundgut}]{vandyk96}
{Van Dyk}, S.~D., Weiler, K.~W., Sramek, R.~A., {et~al.} 1996, AJ, 111, 1271

\bibitem[{Wade \& Horne(1988)}]{wade88}
Wade, R.~A., \& Horne, K. 1988, ApJ, 324, 411

\bibitem[{Wizinowich {et~al.}(2006)Wizinowich, Mignant, Bouchez, Campbell,
  Chin, Contos, van Dam, Hartman, Johansson, Lafon, Lewis, Stomski, Summers,
  Brown, Danforth, Max, \& Pennington}]{wizinowich06}
Wizinowich, P.~L., Mignant, D.~L., Bouchez, A.~H., {et~al.} 2006, PASP, 118,
  297

\end{thebibliography}

\end{document}